


%
%

\input phyzzx

\def\Box{
\vbox{\hrule
\hbox{\vrule\vbox to 2mm{\hbox to 2mm{\hfill}}\vrule}
\hrule}}

\def\a{\alpha}
\def\b{\beta}

\def\d{\delta}
\def\s{\sigma}

\def\hh{\hat{h}_{ab}}


\nopubblock

\titlepage

\rightline{ YITP-95-10 }

\title{\bf Back-Reaction on the Topological Degrees of Freedom
in (2+1)-Dimensional Spacetime}

\author{ Masafumi Seriu}
\address{Inter-University Centre for Astronomy and Astrophysics\break
        Post Bag 4, Ganeshkhind, Pune 411007, India\break
}
\andaddress{
  Yukawa Institute for Theoretical Physics\break
   Kyoto University, Kyoto 606, Japan
   \footnote\star{\rm Present address.}
   }

\vskip 10cm
To appear in {\sl Physical Review D}.
\endpage


\abstract{
We investigate the back-reaction effect of the quantum field on the
topological degrees of freedom in (2+1)-dimensional toroidal universe,
${\cal M} \simeq T^2\times {\bf R}$. Constructing a homogeneous model
of the toroidal universe, we examine explicitly the
back-reaction effect of    the Casimir energy  of a
massless, conformally coupled scalar field, with a conformal vacuum.
   The back-reaction causes an instability
    of the universe: The  torus becomes
     thinner and thinner as it evolves,
     while its total 2-volume (area) becomes
     smaller and smaller.
    The back-reaction
      caused by the Casimir energy can be compared with
     the influence   of the negative cosmological constant:
     Both of them  make the system unstable and
     the torus becomes thinner and thinner
     in shape. On the other hand,
     the Casimir energy is a complicated
   function of  the Teichm\"uller parameters $(\tau^1, \tau^2)$
    causing  highly non-trivial
   dynamical evolutions, while the cosmological constant is
   simply a constant.

   Since the spatial section is a 2-torus, we shall write down the
  partition function of this system,  fixing the path-integral
  measure for gravity modes, with the help of  the  techniques developed
  in string theories. We show explicitly that the partition function
   expressed in terms of the canonical variables
   corresponding to  the (redundantly large) original
  phase space, is  reduced to the partition function defined
  in terms of  the physical-phase-space variables
  with a standard Liouville measure.
  This result is compatible with the general theory of
  the path integral for the 1st-class constrained systems.
}

\endpage


\chapter{Introduction}

Topological considerations are necessary in many situations.
 Since physical laws are usually expressed in terms of local,
 differential equations, their  importance is not prominent at first
  sight. However, once one proceeds to solve them, one has to
  take boundary conditions into account, which allow  the topological
  information to enter in  the theory. In general relativity,
  which handles the dynamics of spacetime, the topological properties
   acquire dynamical meaning and their consideration becomes
  more significant. The aim of this paper is to present
  an  explicit, detailed investigation of the dynamics of topological
  degrees of freedom in spacetime, in the context of the back-reaction
  problem in semiclassical gravity.
   We concentrate on the case of $(2+1)$-dimensional toroidal
   spacetime, ${\cal M} \simeq  T^2 \times  {\bf R}$, and make use of
   various techniques developed for the 2- and 3-dimensional gravity.
  Here, we do not discuss  the topology change [1][14].
  The term ``topological degrees of freedom",
  indicates those  global parameters,
 describing the global deformations of the spatial hypersurface,
 which are of  topological origin (the moduli
 deformations)
 ($\S\S 3-b$ and  Appendix $A$).

 As a first preliminary study for the full quantum gravity,
 it is reasonable to consider the effect of the curvature of
 a fixed background spacetime on the behavior of
 quantum matter field, which is
 the subject of quantum field theory on a curved spacetime [2][3].
 Then, the next natural step is the investigation of the
 influence of such a quantum field on classical spacetime geometry,
 which is called the back-reaction problem in semiclassical gravity.
 Usually, one tries to describe this effect by the semiclassical
  Einstein equation,
$$
G_{\a\b}=\a <T_{\a\b}>\ \ \ ,
\eqno(1)
$$
where $<T_{\a\b}>$ is some $c$-number,  obtained from
 the energy-momentum tensor operator and  the inner-product of
  some quantum states,
 and $\a$ is an appropriate gravitational constant with
 physical dimension $[\a]=[({\rm length})^{n-2}]$. (Here, $n$ is the spacetime
 dimension. We treat $\hbar$ as  $[\hbar]=[1]$, and set $c=1$
  in this paper.)
 There are several uncertain issues and   technically complicated points
  about this treatment.
  First, it is not  clear what kind of quantity  should
  be chosen for $<T_{\a\b}>$ [4]. Here,
  we regard that
  $<T_{\a\b}>$ should be some expectation value,
  rather than the quantity
  $<out|T_{\a\b}|in>$, since the latter
  harms the reality and the causal nature of eq.(1) [5][6][7].
   Then, if one regards the path-integral formalism as fundamental for
     quantum gravity, the so-called
   in-in formalism [8][5][6] should
   be of more importance than the standard in-out
   formalism [7]. Second, the regularization of $<T_{\a\b}>$
   requires complicated, though well-established, techniques,
   which itself is one main topic of the quantum field theory
   on a curved spacetime [2][3]. Third, eq.(1) in general becomes
    complicated, even though  $<T_{\a\b}>$ has been
   successfully computed, so that
   it is difficult to solve it and study the
   effect of the back-reaction in detail.
    Fourth, one can show that eq.(1) can be obtained from the first
   variation of the phase part in the in-in path-integral
   expression [5][7], in which matter part has been
   integrated out formally while gravity part is left unintegrated
    without the explicit  fixation of the measure. If one wants to go
    one step further, however, one should also take care of the
    effect coming from the path-integral measure for the gravity
    part. It is usually difficult since a reasonable, general measure
     has  not been fixed yet.
   Fifth, to speak rigorously,
    eq.(1) itself contains an inconsistency  from the very beginning.
 Since gravity and matter couple, quantum fluctuations of matter
  cause corresponding quantum fluctuations of gravity.
  Thus, there is a limitation
   in principle to  the semiclassical treatment (eq.(1)),
   because  we try to treat gravity classically while matter is
   treated by quantum theory [7]. Specifying the exact
    validity conditions for eq.(1) is one of the main
    topics of semiclassical gravity [7][9][10].

  In this paper, we consider a $(2+1)$-dimensional spacetime
  ${\cal M}\simeq  \Sigma \times {\bf R}$, with $\Sigma \simeq T^2$,
  a torus. We choose, as a matter field, a massless
  conformally coupled scalar field with a conformal vacuum, and
  investigate explicitly  the back-reaction effect, resulting from
  the Casimir energy of matter, on the topological degrees of freedom,
  i.e. the modular-deformations of the torus. As is stated above,
  the topological degrees of freedom is one of the  essential
   ingredients of spacetime dynamics. However, the back-reaction
   on topological modes has seldom been discussed so far, partially
   because such a finite number of degrees of freedom are hidden
   in infinite number of gravity modes in 4-dimensional spacetime.
    One advantage of the reduction of the number of dimension
    from 4 to 3 is that, only the  finite topological modes plus a
    spatial volume remain dynamical for the case of pure gravity,
     due to the dimensionality [11][12][13].
    One can understand  this point as follows: When $n=3$,
     the spatial metric $h_{ab}$ has 3 independent
   components at each spatial point, while there are 3 constraints
   at each point. Thus, redundant infinite number of modes are gauged
    away and only a finite number of modes remains.
     Here, we want to investigate the back-reaction effect from
     matter onto the topological degrees of freedom of spacetime,
      which would force us to  take the matter field
     into account. To preserve the above-mentioned
     nice property of the finiteness
     of the number of degrees of freedom, we choose a model in which
     the matter field is in a vacuum state on a  spatially homogeneous
     (2+1)-dimensional spacetime.
 Another advantage of the reduction of  dimension in   the discussion
 of topological aspects comes from  the fact that
  2-dimensional topology is completely classified in a simple manner
  so  that
 it is easy to construct various topologies [14].

 Another good point of this model is that some difficulties and
 complications stated above
 of the semiclassical Einstein equation, eq.(1), become
 simplified and tractable to a great extent in this case:

 First, we choose a conformal vacuum $|0>$ as a
 natural
 candidate for a vacuum state of matter in our case,
 and use
 $<0|T_{\a\b}|0>$ on the right-hand side of eq.(1).

 Second, since the
 background
 spacetime shall  be chosen as
 (conformally) flat and the matter field is conformally invariant,
   $<T_{\a\b}(g)>$ can be calculated from
 $<T_{\a\b}(\eta)>$ ($\eta$: a flat metric) along with  the
 trace-anomaly [2], which  which simplifies the manipulation. Furthermore,
 in our case, the spacetime dimension is odd, $n=3$, so that  there
  is no trace-anomaly [2]. Thus,  $<T_{\a\b}(g)>$ is
  related to $<T_{\a\b}(\eta)>$ in a simple manner.

 Third, because  of the dimensionality,
 eq.(1) is reduced to a set of
 six first-order ordinary differential equations and we can
 investigate the effect of the back-reaction explicitly.

 Fourth, we restrict the metrics
  to a special class,  with
 spatial part  being the one for the  locally flat metrics on a torus.
  Thus, we can fix the path-integral measure  by the
  use of the techniques developed in string theories [15][16].
 Within this model, we can discuss explicitly the
   influence on the semiclassical dynamics of gravity.
   Our treatment corresponds to the minisuperspace approach in quantum
   cosmology: Putting  restrictions on the variables to be quantized
   (e.g. spatial homogeneity),
   which is compatible with the classical dynamics,  quantum
   theory is to be constructed
   within this restricted subclass of variables. Though this treatment
   is self-consistent as a quantum system, one significant question
   naturally arises: To what extent such a treatment reflects faithfully
   the original full quantum theory? From the viewpoint of the original
   full system, the restrictions are regarded as constraints on
   the phase space, which can modify the path integral measure for
   the reduced variables (minisuperspace variables). Our model may be
   a good test candidate to investigate this point in detail.

   Fifth,  the (in-in) effective action for
   gravity, $W[g_+:g_-]$, becomes relatively simple in our case,
   and  this reduces to
   $W[ \tau^1_+, \tau^2_+, V_+: \tau^1_-, \tau^2_-, V_-]$,
   a functional of six functions of $t$,
   $( \tau^1_\pm, \tau^2_\pm, V_\pm)$, where $V_\pm$ indicate
   the spatial 2-volume (area) and  $(\tau^1_\pm, \tau^2_\pm)$
    are the Teichm\"uller parameters describing the topological
    degrees of freedom of a torus. Although the exact calculation of
  $W$ has already become difficult, we can still
  estimate its functional form
  to leading order in $\hbar$. In computing $W$,
  our model reveals explicitly the
  peculiarity of the semiclassical gravity, compared with the
  standard treatment of the quantum dissipative system, e.g. the
  Brownian motion [17]: There is no linear coupling between
  the sub-system (gravity) and the environment (matter field). Their
   coupling is put in the kinetic term of the matter field.
 This model might provide  the simplest non-trivial example
 for the investigation of
 the quantum dissipative system including gravity.

 In \S2, we recapitulate how to handle quantum fields on
 topologically non-trivial spaces: Construct the quantum field theory on
 ${\cal M} \simeq  T^2 \times {\bf R}  $ and calculate the
 Casimir energy of a massless, conformally coupled scalar
 field with a conformal vacuum.[2],[3],[19]

 In \S3, we extract explicitly the  topological degrees of freedom
 of a torus and reduce eq.(1) to a canonical system with a finite
 number of degrees of freedom [13]. Then, we investigate explicitly
 the effect of the back-reaction of matter on the dynamics of the
 topological degrees of freedom. We shall see that the back-reaction
 makes the system unstable and the torus becomes
 thinner and thinner as it evolves,
 while its 2-volume becomes smaller and smaller.
 These behaviors are   universal that is independent of
 the initial conditions. The asymptotic analysis of the set of
 dynamical equations justifies this point. We shall also compare
 our case of the Casimir energy
 with the case of the negative cosmological constant,
 since both of them  can be regarded as    negative energies. Most
 significant  difference is that   the Casimir energy is a complicated
 function of the Teichm\"uller parameters $(\tau^1, \tau^2)$, while
 the negative cosmological constant is just a constant.

 In \S4, we investigate  the partition function of this system, fixing
 the measure  with the help of the techniques
 in string theories.
 We show explicitly that the gauge-fixing reduces
 the partition function formally
   expressed in terms of the canonical
   variables for the (redundantly large) original
  phase space,  to the partition function defined
  in terms of  the physical-phase-space variables
  with a standard Liouville measure.
  This result is compatible with the general theory of
  the path integral for the 1st-class constrained systems.
 We also  estimate the functional form of $W$
  to  leading order in $\hbar$. Section 5 is reserved for discussions.


\chapter{Quantum field theory on a (2+1)-dimensional toroidal spacetime}

This section is for defining the model to be considered, and
calculating the energy-momentum tensor in our model, as a
preliminary for the next section, where the back-reaction effect is
analyzed in detail. Calculating $<T_{\a \b}>$ is now a well-established
topic, and we just sketch the essence in the context of our model for
later uses.
\item{(a)} {\bf  Scalar field on a torus}

We consider a (2+1)-dimensional spacetime with  topology
$ T^2 \times {\bf R} $. We concentrate on the case when the  geometry
of the space $\Sigma \simeq T^2$ is locally flat.
  A  flat 2-geometry is endowed on $\Sigma$ by giving
 a metric
\footnote\star{
    For definiteness, we shall use the symbol ${\hat h}_{ab}$
    to represent the particular matrix given by $(2-b)$, while
    the symbol $h_{ab}$ shall be reserved  for more general context,
    representing a general spatial metric induced on a spatial
     surface $\Sigma$.
}
 ,
$$
dl^2= \hh d\xi^a d\xi^b\ \ \ ,
\eqno(2-a)
$$
where
$$
\hh  ={ 1\over  \tau^2 }
\pmatrix{ 1     &  \tau^1  \cr
         \tau^1 &  |\tau|^2 \cr}\ \ \ ,
\eqno(2-b)
$$
and the periodicities in the coordinates $\xi^1$ and $\xi^2$ with
period 1 are understood.
Here
\footnote\S{
   Throughout this paper, $\tau^2$
   always indicates  the second component of $(\tau^1,\tau^2)$, and not
   the square of $\tau$. The latter  never appears in the formulae.
}
, $(\tau^1, \tau^2)$ are the Teichm\"uller parameters [15][16]
independent of spatial
coordinates $(\xi^1, \xi^2)$, and $\tau := \tau^1 + i \tau^2$,
$\tau^2 >0$.
 Note that $ \sqrt{\hat h}:=({\rm det} \hh )^{1/2} =1$.

  The Laplacian operator
  $\Delta :=-1/ \sqrt{h}.\partial_a(h^{ab} \sqrt{h} \partial_b)$
  on $\Sigma$ with the line element $dl^2$
 (eq.$(2-a,b)$) gives
the  normalized eigenfunctions
 $$
f_{n_1 n_2} (\xi) = \exp (i2\pi n_1 \xi^1)\cdot  \exp (i2\pi n_2 \xi^2)
 \ \ \ (n_1, n_2  \in  {\bf Z})\ \ \ ,
 \eqno(4)
 $$
and the  eigenvalues
$$
\lambda_{n_1 n_2}={4\pi^2 \over \tau^2}
 (|\tau|^2 n_1^2
             -2\tau^1 n_1 n_2 + n_2^2)\ \ \ .
\eqno(5)
$$

 Now, let us consider a spacetime
 ${\cal M} \simeq \Sigma \times {\bf R} $,
 with a line element $ds^2=-dt^2+ \hh d\xi^a d\xi^b$. The fundamental
 positive frequency  solutions for  $\Box u(t,\xi^1,\xi^2)=0$ are
\footnote\star{
In connection with the later applications, it is worthwhile to note  that,
 even though $\tau$ would depend on $t$,   the form of
the  equation $\Box \psi =0$ would not change,
because of the form of the metric,
 $g_{\alpha \beta}=(-1, \hh )$ with $det g_{\alpha \beta}=-1$.
}
 $$
 {\overline u}_A(t,\xi)={1\over \sqrt{2\omega _A}}
 e ^{-i\omega _At}f_A(\xi)\ \ \ ,
 \eqno(6)
 $$
 where $A$ stands for $n_1 n_2$ and
 $\omega _A:= \sqrt{\lambda_A}=\sqrt{\lambda_{n_1 n_2}}$.
 Afterwards,
  we follow the standard procedure for the field quantization
[2][3].


\item{(b)} {\bf  The model}

 We shall investigate the back-reaction of the matter field on the
 topological degrees of freedom $(\tau^1, \tau^2)$. The most ideal
  treatment of the back-reaction described by eq.(1) may be
 the self-consistent determination of the geometry $g_{\a\b}$
  through eq.(1): $<T_{\a\b}>$ depends on $g_{\a\b}$, and this
  $g_{\a\b}$ is self-consistently determined by eq.(1).
   However, it turns out that such a treatment becomes highly
 complicated even in our simple model. To make our analysis tractable,
 then, we treat the back-reaction in the following sense, which
 is usually adopted in the  back-reaction problems [2][3]:
 We prepare a background
 spacetime and calculate $<T_{\a\b}>$ on it. Then, we discuss
  the  modification of the background geometry
  due to the $<T_{\a\b}>$, using eq.(1).

Now, as a background spacetime,
we choose a solution of the vacuum Einstein equation, $G_{\a\b}=0$.
More specifically,
we prepare a locally flat spacetime,
$ds^2=-dt'^2 + V dl^2
 =V (-dt^2+dl^2)$,
  where $dl^2$ is given by eqs.$(2-a,b)$, and
  $V$, $\tau^1$ and  $\tau^2$ are chosen to be constant for the background
  spacetime. (Below, we occasionally treat this flat spacetime as
  conformally flat, just for  mathematical convenience.)
    We choose as a matter
   field, a massless conformally coupled scalar field $\psi$,
$$
S_m=-{1\over 2} \int (g^{\alpha \beta}\partial _\alpha \psi
        \partial _\beta \psi
              + {1\over 8}R \psi^2)\sqrt{-g} d^3x\ \ \ .
\eqno(7)
$$
The (improved) energy-momentum tensor operator [3] becomes,
$$
T_{\alpha \beta}(g)={3\over 4}\partial_\alpha \psi \partial_\beta \psi
 -{1\over  4}\partial_\gamma \psi \partial^\gamma \psi g_{\alpha \beta}
 -{1\over 4}\psi \partial_\alpha  \partial_\beta \psi
 +{1\over 12}\psi \Box \psi g_{\alpha \beta}
 +{1\over 8} \psi^2 (R_{\alpha \beta}-{1\over 3}g_{\alpha \beta}R).
\eqno(8)
$$
 We choose the conformal vacuum as a vacuum state for the matter field.
 Then, $<T_{\alpha \beta}(g)>$ is simply
 related to $<T_{\alpha \beta}(\eta)>$ as,
 $$
 <T_{\alpha \beta}(g)>= V ^{-1/2}<T_{\alpha \beta}(\eta)>\ \ \ ,
 \eqno(9)
$$
when
the metric $g_{\a \b}$ and the flat metric $\eta_{\a \b}$ are
related as $g_{\a \b}=V \eta_{\a \b}$.
\footnote\star{
This simplification occurs, because $<T_{\alpha \beta}(g)>$ for a
conformally invariant field, with the conformal vacuum, on
a conformally flat spacetime is completely determined by
$<T_{\alpha \beta}(\eta)>$
 and the trace anomaly $<T_\alpha ^\alpha (g)>$, while the
 latter vanishes when the spacetime dimension is odd [2]. }
On  flat spacetime, the
field equation for $\psi$ becomes $\Box \psi =0$, and
eq.(6) can be used as fundamental solutions.
In this manner,
the time evolution of $V$ causes no direct complication in the
analysis.

However, the time-dependence of $(\tau^1, \tau^2)$
 caused by the back-reaction harms the self-consistency of the
 analysis, which is inevitable if the tractability of the
 back-reaction problem, described by eq.(1), is to be maintained.
  When $(\tau^1, \tau^2)$
 evolve in time, the functions in eq.(6) are no longer exact solutions
 for $\Box \psi =0 $, because
 $\omega _A:= \sqrt{\lambda_A}$ becomes $t$-dependent, through
  the $t$-dependence of $(\tau^1, \tau^2)$ (eq.(7)). Furthermore,
  the spacetime described by
  $ds^2=-dt^2 + V(t) \hh d\xi^a d\xi^b$ is no longer conformally
  flat when $(\tau^1, \tau^2)$ evolves, because of the $t$-dependence
   of $\hh$.
   Thus, we should look at the results of the analysis in an adiabatic
   sense, i.e. valid when terms including
    $\dot \tau^1$ and   $\dot \tau^2$
   are not dominant in the formulae prominently. Such a conflict between
   self-consistency and the tractability of the analysis always
    occurs in the back-reaction problem.
    In our present model, this adiabatic treatment
    provides a good approximation  because
    $\dot \tau^1$ and   $\dot \tau^2$,  caused by the back-reaction,
    turn out to be sufficiently small (see $\S\S 3-c$).

 We next need   Hadamard's elementary function [2][3]
 $G^{(1)}(x)$ for $ds^2 = -dt^2 + dl^2$ to  calculate
  $<T_{\alpha \beta}(\eta)>$.
  This function  and the related energy-momentum tensor have already
  been extensively investigated [19].
 We first compute $G^{(1)}(x)$
 for ${\cal M} \simeq {\bf R}^3$, and afterwards take care of the
 periodicity in ${\cal M} \simeq  T^2 \times  {\bf R}$,
 adding all contributions from points which should be
 identified [18][19]. For  the 3-dimensional Minkowski space,
  $G^{(1)}(x)$ is,
$$
G^{(1)}(x):=<0|\{\psi (x),\ \psi (y) \}|0>
={\hbar \over 2\pi} (2\sigma)^{-1/2} \ \ \ (\sigma >0)\ \ \ ,
\eqno(10)
$$
where $\sigma :={1\over 2}x^2 ={1\over 2}\eta_{\a\b}x^\a x^\b$,
${1\over 2}$ times a square of a world distance. Thus we get
\footnote\star{
The prime attached to the
$\Sigma$-symbol, like in eq.(11),  indicates
that the zero-mode ($n_1=n_2=0$)  should be excluded from the summation,
whenever it causes a divergence.}
$$
G^{(1)}(x)={\hbar \over 2\pi}
\sum_{n_1,n_2=-\infty} ^{\infty\ \ \prime }(2\sigma_{n_1n_2} (x))^{-1/2}\ \ \ ,
\eqno(11)
$$
where
$$
2\sigma_{n_1n_2} (x):=-t^2+  {1\over \tau^2}
      \left|(\xi^1 + n_1)+\tau (\xi^2 + n_2)   \right|^2  \ \ \  .
$$
Now it is straightforward to compute $<T_{\alpha \beta}(\eta)>$
 explicitly.
\footnote\S{
 For computation it is helpful to note that
$
 <T_{\alpha \beta}(\eta)>
    ={1\over 2}\partial_\alpha \partial_{\beta'}G_{\sim}^{(1)}
$,
where $\partial_\alpha \partial_{\beta'}G_{\sim}^{(1)}
 :=\partial_{x^\alpha} \partial_{{x'}^{\beta}}
 G^{(1)}(x-x')_{|x'\rightarrow x} $ and
 $x^{\alpha}:=(t, \xi^1, \xi^2)$.
 } ($\eta_{\a \b}=(-1, \hh)$ with $(2-b)$.) The result is,
$$
<T_{00}>=- {\hbar {(\tau^2)^{3/2}} \over {4\pi}}
                 \sum_{n_1n_2}^{\ \ \ {}_\prime}
         {1\over |n_1+\tau n_2|^3}\ \ \ ,
\eqno(12-a)
$$
$$
<T_{11}>={\hbar {(\tau^2)^{1/2}}\over {4\pi}}\sum_{n_1n_2}^{\ \ \ {}_\prime}
                       {1\over |n_1+\tau n_2|^3}
   -{3 \hbar {(\tau^2)^{1/2}}\over {4\pi}}
\sum_{n_1n_2}^{\ \ \ {}_\prime}
         {(n_1 + \tau^{1} n_2)^2 \over  |n_1+\tau n_2|^5 }\ \ \ ,
\eqno(12-b)
$$
$$
<T_{22}>={\hbar (\tau^2)^{1/2}|\tau|^2 \over 4\pi}
        \sum_{n_1n_2}^{\ \ \ {}_\prime} {1\over |n_1+\tau n_2|^3}
   -{3 \hbar ({\tau^2)^{1/2}} \over {4\pi}}\sum_{n_1n_2}^{\ \ \ {}_\prime}
{(\tau^{1}n_1+|\tau|^2 n_2)^2 \over  |n_1+\tau n_2|^5 }\ \ \ ,
\eqno(12-c)
$$
$$
\eqalign{
<&T_{12}>=<T_{21}>\cr
    &={\hbar { \tau^1(\tau^2)^{1/2}} \over {4 \pi}}
    \sum_{n_1n_2}^{\ \ \ {}_\prime}
               {1\over |n_1+\tau n_2|^3}
   -{{3\hbar  (\tau^2)^{1/2}}  \over {4\pi}}\sum_{n_1n_2}^{\ \ \ {}_\prime}
{{(n_1+\tau^1 n_2)(\tau^{1}n_1+|\tau|^2 n_2)  } \over  |n_1+\tau n_2|^5 }
                ,                            \cr
}
\eqno(12-d)
$$
$$
<T_{0a}>=<T_{a0}>=0\ \ \ (a=1,2) \ \ \ .
\eqno(12-e)
$$
 For a metric $g_{\alpha \beta}= V (-1, \hh)$,
$<T_{\a\b}(g)>$ can be obtained by eq.(9).
Since the Planck scale is the only scale which comes into
our model, we understand that a suitable power of
$\alpha:=l_{\rm Planck}$ is multiplied to quantities like those in
eqs.$(12-a)-(12-e)$, if necessary, in order to adjust their
physical dimensions.
These  contributions of order $\hbar$ to $<T_{\alpha \beta}>$ in
eqs.$(12-a)-(12-e)$  originate from a non-trivial spatial topology
$\Sigma \simeq T^2$, and are well-known as the Casimir effect [2][3].


\chapter{Back-reaction of the Casimir effect on the topological
degrees of freedom}


\item{(a)} {\bf  The extraction of dynamics of the modular deformations}

 Having computed $<T_{\alpha \beta}(g)>$ in the previous section,
 we shall now investigate the back-reaction of $<T_{\alpha \beta}(g)>$
  on the evolution of the spacetime. We consider the Einstein
  gravity on  ${\cal M}\simeq  T^2  \times {\bf R}$ and a  massless
   conformally coupled scalar field on it;
 $S= {1\over \alpha} \int R \sqrt{-g} + S_m$,
 where $\alpha:=l_{\rm Planck}$ and $S_m$ is given by eq. (7).
 The canonical formulation is suitable to investigate the
 temporal evolution of the spacetime. We thus perform the
 (2+1)-decomposition, but  care should be taken because of
  the presence of
 the conformally coupled field.
  In the back-reaction problem,  we regard that $\psi^2(x)$ is
  replaced by a vacuum expectation value $<\psi^2(x)>$, which
   is  independent of spatial coordinates. Furthermore, we shall
   finally choose the spatial coordinates s.t. $N^a =0$ so that
   $n^a = (- 1/N, {\vec 0})$. These facts simplify the procedure
   of $(2+1)$-decomposition.

 Following the standard
 manipulation [20], we finally get the total action
 in canonical form,
$$
S=\int \pi^{ab} \dot{h_{ab}} -N{\cal H} -N^a{\cal H}_a\ \ \ ,
\eqno(13-a)
$$
where the Hamiltonian constraint
and the momentum constraint  become,
respectively,
$$
\eqalignno{
{\cal H}&=\{(K_{ab}K^{ab}-K^2- {}^{(2)}  \!  R)/\a + <T_{\a\b}>n^\a n^\b \}
  \sqrt{h}\ \ \ ,
                                 &(13-b)   \cr
{\cal H}_a/\sqrt{h}&=-2D_b({K_a}^b -{\d_a}^bK)/\a-<T_{a\b}>n^\b\ \ \ .
                                 &(13-c)   \cr
}
$$
Here, $N, \ N_a$ are the lapse and the shift functions,
$n^\a =\left(-1/N, N^a/N   \right)$ is the normal unit vector
 of the spatial surface, and $^{(2)}  \! R$ stands for the scalar curvature
 for the spatial surface $\Sigma$.
 The operator $ D_a$ is the covariant derivative
  w.r.t.  $ h_{ab}$,  and
$\pi^{ab}:=(K^{ab}-Kh^{ab})\sqrt{h} /\a$, $K_{ab}$ is the extrinsic
   curvature of a  spatial surface.
\footnote\star{
    Throughout this paper,
    we use a spatial metric $h_{ab}$, an induced metric on a
    spatial surface $\Sigma$, to raise and lower the spatial indices,
    $a, b, c, \cdots$, and to define the spatial covariant derivative
    $D_a$.
    In particular, the geometry of our concern is given by the line
    element, $ds^2 = -dt^2 + V dl^2$, with $(2-a,b)$. Thus, the spatial
    metric in our model is $h_{ab}= V \hh$, with  $(2-b)$.
}

 We choose a  coordinate system s.t. $N^a=0$ so that
$ n^\a=( -1/N, 0)$. Thus, $<T_{a\b}>n^\b = -1/N.<T_{a0}>=0$ ($a=1,2$)
from eq.$(12-e)$. In our case, thus, the momentum constraint becomes
$$
{\cal H}_a/\sqrt{h}=-2D_b({K_a}^b -{\d_a}^bK)/\a=0\ \ \ .
\eqno(13-c')
$$
Then, we can extract the moduli degrees of freedom (corresponding to
the global deformations of a torus) by
 solving eq.$(13-c')$ explicitly [13].

The system of coordinates in our model
($ds^2=-dt^2+Vdl^2$ with $(2-a,b)$) corresponds to
 York's time-slicing [21],
 i.e. the time-slicing by the  spatial surfaces on which
 $$
 \s :=-K/\a ={\rm const}\ \ \ .
 \eqno(14)
 $$
 Thus, eq.$(13-c')$ is equivalent to
$$
\a {\cal H}_a/\sqrt{h}=-2D_b{{\tilde K}_a}^b=0 \ \ \ ,
\eqno(13-c'')
$$
where  ${{\tilde K}_a}^b:={K_a}^b -{1\over 2}{\d_a}^bK$,
the traceless part of ${K_a}^b$. It means that
\footnote\star{
See Appendix $A$ for the terminology and notations related to the
moduli space.}
${\tilde K}^{ab} \in  Ker{P_1}^\dagger$, so that
 ${\tilde K}^{ab}$ can be expanded in terms of the basis of
$Ker {P_1}^\dagger$, $\{ {\Psi ^A}^{ab} \}_{A=1,2}$ ;
$$
{\tilde K}^{ab}= {1\over \a} \sum_{A=1}^2 p_A {\Psi ^A}^{ab}\ \ \ .
\eqno(15)
$$
In our case, $\Sigma \simeq T^2$, we can choose the lapse
function $N$ as $N=N(t)$ without any contradiction with the York's
slice. This is shown almost in the same manner as for the case of
pure $(2+1)$-gravity [12][13].
 Now, using some basic facts on the moduli space ${\cal M}_{g=1}$
 (see Appendix $A$),  it is straightforward [13] to show that
  our system is  reduced to
 $$
\eqalignno{
 S=\int dt& \s {dV\over dt} + \sum_{A=1}^2 p_A {d\tau ^A\over dt} \cr
   & - {N(t)\over \a}(\sum_{A,B=1}^2{\cal G}^{AB}p_Ap_B
        -{1\over 2}\a^2\s^2V +\a <T_{\a\b}>n^\a n^\b V ) .&(16)\cr
}
$$
Note that the contribution from the spatial diffeomorphism
 has been eliminated from dynamics, by solving the momentum
 constraint $(13-c')$ explicitly. Only the
 Weyl deformations and the modular deformations have remained.


\item{(b)} {\bf  The evolution of the Teichm\"uller
parameters caused by the back-reaction}

In our model, $ds^2=V(-dt'^2 + \hh d\xi^a d\xi^b)$. Thus, from
 eq.(9) and $n^\a =(-1/\sqrt{V}, 0)$, we get
 $<T_{\a\b}>n^\a n^\b = V^{-3/2} <T_{00}>$, where $<T_{00}>$ is given
 by eq.$(12-a)$. (Note that this combination is coordinate independent.)

 By setting $N(t)=1$, we get the canonical equations of motion described
 by the constraint  function,
 $$
 \a H =\sum_{A,B=1}^{2}{\cal G}^{AB}p_Ap_B
   -{1\over 2}\a^2 \s^2 V - \hbar \a (\tau^2)^{3/2}f(\tau) V^{-1/2} =0\ \ \ ,
 \eqno(17)
 $$
 where
$$
\eqalignno{
 f(\tau^1, \tau^2)
     :&={1\over 4\pi}\sum_{n_1,n_2=-\infty}^{\infty_{{}_\prime}}
                    {1\over |n_1+\tau n_2|^3} \cr
     &={1\over 4\pi}\sum_{n_1,n_2=-\infty}^{\infty_{{}_\prime}}
   {1\over (n_1^2+2\tau^1n_1 n_2 +|\tau|^2n_2^2)^{3/2}}\ \ \ .& (18) \cr
}
$$
Clearly, $f(-\tau^1, \tau^2)=f(\tau^1, \tau^2)$,
$f(\tau^1+n, \tau^2)=f(\tau^1, \tau^2)$,
$f(n+a, \tau^2) = f(n-a, \tau^2)$ ($n$: integer, $a$: real)
and $f(\tau^1, \tau^2)$ is singular at $(\tau^1, \tau^2)=(n,0)$.
Furthermore, the combination
$2\pi (\tau^2)^{3/2}f(\tau^1, \tau^2)$ appearing in (17)
is equivalent to
the non-holomorphic
Eisenstein series $G(\tau, 3/2)$,
whose modular invariance as well as
 other properties are well-known [22]. The first term in eq.(17) is
 also modular invariant, since it behaves as a scalar field on the
 moduli space.
\footnote\star{
    Another convenient way for  discussing the invariance is to  perform
    the Legendre transformation of the action
    in concern,  and to look at the action in terms of the  configuration
    variables. In this case, the kinetic term for $(\tau^1, \tau^2)$
    becomes  proportional to
    $\sum {\cal G}_{AB}\dot{\tau}^A \dot{\tau}^B $, which
    is clearly modular invariant. For the discussion in the context
     of the  path-integral, including the discussion on the
     path-integral measure, see $\S\S 4-a$.
 }
 Thus, the Hamiltonian constraint eq.(17) is modular invariant as
 it should be.
 Figures $1-a,b$ show the behavior of the function $f(\tau^1, \tau^2)$.

 For the explicit investigation of the dynamics, let us first
 calculate ${\cal G}^{AB}$ according to eqs.$(A5)$ and $(A2-c)$ with
 $h_{ab}={V\over {\a^2 \tau^2}}
   \pmatrix { 1      &   \tau^1       \cr
              \tau^1 &   |\tau|^2      \cr}$.   (Note that
 $ds^2=-dt^2 + V dl^2$.)
Then, we get
$$
{\cal T}_{1ab}={V \over { \a^2 \tau^2}}
   \pmatrix { 0      &  1       \cr
                            1      &  2\tau^1      \cr},\ \
{\cal T}_{2ab}={V \over { \a^2 (\tau^2)^2}}
   \pmatrix { -1           &  -\tau^1                    \cr
              -\tau^1      &  (\tau^2)^2-(\tau^1)^2      \cr}\ \ .
\eqno(19-a)
$$
Note that
$\{ {\cal T}_{Aab} \}_{A=1,2}$ are symmetric, traceless 2-tensors
 satisfying $-2D_b{{\cal T}_{Aa}}^b
 =-2\partial_b {{\cal T}_{Aa}}^b=0$. Thus, $\{ {\cal T}_{Aab} \}_{A=1,2}$
 can also be utilized to form a basis for $Ker {P_1}^{\dagger}$,
 $\{  {\Psi^A}^{ab} \}_{A=1,2}$. By normalizing them to satisfy
$( \Psi ^A, {\cal T}_B) ={\d^A}_B$, we obtain,
$$
\Psi^1_{ab}={1 \over 2 }
   \pmatrix { 0      &  \tau^2       \cr
              \tau^2      &  2\tau^1 \tau^2     \cr},\ \
\Psi^2_{ab}={ 1 \over 2 }
   \pmatrix { -1           &  -\tau^1                    \cr
              -\tau^1      &  (\tau^2)^2-(\tau^1)^2      \cr}\ \ .
\eqno(19-b)
$$
 Thus, the Weil-Peterson metric reduces to the one
 which is conformally equivalent  to the Poincar\'e metric,
 $$
\eqalign{
{\cal G}_{AB} &=({\cal T}_A, {\cal T}_B)
              ={2 V \over {\a^2 (\tau^2)^2} }
                \pmatrix{1 & 0 \cr
                         0 & 1 \cr }         \ \ \ , {} \cr
{\cal G}^{AB} &=( \Psi^A, \Psi^B)
               = {\a^2 (\tau^2)^2 \over {2V}}
                \pmatrix{1 & 0 \cr
                         0 & 1 \cr }      \ \ \ . \cr
}
\eqno(20)
$$
 Hence, the geometry conformal to
the Poincar\'e geometry [23] (negative constant curvature geometry)
is endowed on the Teichm\"uller space,
which is equivalent to the upper half-plane $H_+$
($(\tau^1, \tau^2)\in {\bf R}\times {\bf R}_+$ ).
Then, the system has been finally reduced to the constraint system
 $\left( (V,\s), (\tau^1, p_1), (\tau^2, p_2); H=0  \right)$ with
$$
\a H= { \a^2 (\tau^2)^2 \over {2V} }
 (p_1^2 + p_2^2)-{1\over 2}\a^2 \s^2 V
        - \hbar \a (\tau^2)^{3/2}f(\tau) V^{-1/2}=0\ \ \ .
\eqno(21)
$$
 The equations of motion for $(V, \s)$ are
$$
\eqalignno{
{\dot V}&=-\a \s V & (22-a) \cr
{\dot \s}&={\a \over 2} \s^2 + {\a (\tau^2)^2 \over {2V^2}}(p_1^2+ p_2^2)
    - {\hbar \over 2} (\tau^2)^{3/2} f(\tau) V^{-3/2}\ \ \ . & (22-b) \cr
}
$$
 The equations of motion for $(\tau^1, p_1)$ and $(\tau^2, p_2)$ are
 $$
 \eqalignno{
 \dot{\tau}^1 &={\a \over V} (\tau^2)^2 p_1 \ \ \ , & (23-a) \cr
 \dot{p_1} &=\hbar (\tau^2)^{3/2}
           { {\partial f(\tau)} \over {\partial \tau^1}}
          V^{-1/2} \ \ \ , & (23-b) \cr
 \dot{\tau}^2 &={\a \over V} (\tau^2)^2 p_2 \ \ \ ,& (24-a) \cr
 \dot{p_2} &= -{\a \over V} \tau^2 (p_1^2 +p_2^2)
              +{3\hbar \over 2}(\tau ^2)^{1/2} f(\tau) V^{-1/2}
              +\hbar (\tau^2)^{3/2}
               { {\partial f(\tau)} \over {\partial \tau^2}}
               V^{-1/2} & (24-b) \cr
}
$$
   First, we  should note that the time evolution becomes trivial
 when there is no matter field, $f(\tau) \equiv 0$, in the
 following sense: In this case, eqs.$(22-a,b)$ allow a solution,
  $\s \equiv 0$, $ V={\rm const}$, $p_1=p_2 \equiv 0$. It is clear
  that, from
  eqs.(21), $(23-a,b)$ and $(24-a,b)$, equations of motion
   do not allow any solution, compatible with
   $\s \equiv 0,\ \ V={\rm const}$,
   other than $\tau^1=$const, $\tau^2=$const. This  corresponds to the
   3-dimensional Minkowski space in the standard coordinates
   $(T, X^1, X^2)$ with suitable
   identifications in spatial section $( X^1, X^2)$ described by
    $(\tau^1, \tau^2)$. The  unique solution above shows that
 there is no time evolution with respect to the standard
time-slice, $T=$const ($\s=0$).  This configuration
is what we have chosen as a background spacetime.
 (However, there are different solutions characterized by the initial
 condition  $\s \neq 0$.
 In these cases, $(\tau^1, \tau^2)$ evolve in time.)

 The back-reaction of the quantum field causes a non-trivial
 evolution of $(\tau^1, \tau^2)$, i.e. global deformations of
a torus. It is clear from eq.(21) that even when $\s \approx 0$
 so that the term $-{1\over 2}\a^2 \s^2 V$ in eq.(21) can be neglected,
  a non-trivial evolution of $(\tau^1, \tau^2)$  occurs,
 because  of the negativity of the term
 $-\hbar \a (\tau^2)^{3/2}f(\tau) V^{-1/2}$ in eq.(21). The choice of
 the solution $\s \equiv 0,\ \ V\equiv {\rm const}$ is not allowed any
 more, as is seen from eqs.$(22-a,b)$.

Figures $2-a,b,c$ show a typical  example of the evolution
of $(\tau^1, \tau^2)$, $(p_1, p_2)$ and $(V,\sigma)$, respectively.
 Units,  s.t. $\hbar =1$ and $\a=1$,  have been chosen.
   We have set  the initial conditions for
  $(\tau^1, \tau^2)$, $p_1$, $\s$ and $ V$. The initial
  condition for $p_2$ has been decided using the constraint
   equation eq.(21).
    We  can observe the same asymptotic behavior of the system
    which arises  irrespective  of the initial conditions, due to the
   back-reaction: The back-reaction drives the system into the
   direction corresponding to a thinner torus, i.e.
   $\tau^2 \rightarrow 0$  while $\tau^1 \rightarrow {\rm finite}$.
   At the same time, the 2-volume $V$ asymptotically approaches
   zero. We find out that this  behavior is universal
   by setting various generic
    initial conditions. This universal behavior can also
    be understood by
    investigating  the qualitative characteristics of eqs.(21)-(24),
    which shall be done in the next sub-section.

    We should also note a special class of trajectories characterized
    by the initial condition $\tau^1 = n/2$ ($n:$ integer),
    $p_1=0$.
    \footnote\star{
    Because of the modular invariance of the system, the cases of
    $\tau^1$=integer are equivalent to the case of $\tau^1=0$, and
    those of $\tau^1$=half integer are equivalent to the case of
    $\tau^1=1/2$. The trajectories of the former cases are stable
    against perturbations, while the trajectories of the latter cases
    are unstable. This can be seen from Figures $1-a,b$
    along with eq.(17).
    }
     The $(\tau^1, \tau^2)$-trajectory becomes
    parallel to the $\tau^2$-axis and $(p_1, p_2)$-trajectory
    is on the $p_2$-axis.
     Depending on whether $p_2 > 0$ or $p_2 > 0$,
    $\tau^2$ tends to $\infty$ or $0$, respectively. In any case,
    the shape of the torus becomes thinner and thinner as it evolves.
    (Note the modular invariance of the system.)


\item{(c)}{\bf The asymptotic behavior of the system}

We can understand
 the  universal behavior of the system by
  looking at eqs.(21)-(24), and investigating the asymptotic behavior
  of the system as $t \rightarrow \infty $. Key behaviors are
\item{(i)}  $V\rightarrow 0$, $\s \rightarrow \infty$,
             $\s V^n \rightarrow \infty$ ($n=1,2,3,\cdots$).
\item{(ii)}
  $(\tau^2)^2(p_1^2 + p_2 ^2)$ increases,
   at least as fast  as
     \footnote\S{
Here, `$y(t)$ increases at least as fast as $x(t)$' or
`$y(t)$ increases at least as  $x(t)$' means that,
$\left| x(t)/y(t)\right| \rightarrow c$, $0 \leq c< \infty$ when
$t \rightarrow \infty$. In other words, $1/y(t) = O(1/x(t))$ when
$t \rightarrow \infty$. }
 $\s^2 V^2$.
\item{(iii)}  $\tau^2 \downarrow 0$, or $\tau^2 \rightarrow \infty$.
\item{(iv)}   $p_1 \dot{\tau}^1 + p_2 \dot{\tau}^2$  increases
  at least as $\s^2 V$, and
  ${1 \over (\tau^2)^2}$$((\dot{\tau}^1)^2 +(\dot{\tau}^2)^2 )$
    increases at least as $\s^2 $.

Now, let us derive the above results.
 First of all,
  eq.$(22-b)$ can be written with the help of eq.(21) as
$$
{\dot \s}= \a  \s^2 +
 {\hbar \over 2} (\tau^2)^{3/2} f(\tau) V^{-3/2}\ \ \ .
\eqno{(22-b')}
$$
 Thus, ${\dot \s}>0$, so that $\s$ always increases and becomes positive
  at some stage. Then, $V$ decreases because of eq.$(22-a)$.
  Furthermore, it is easily
  shown that
  $(\s^2 V^2 \dot{)\ }$$= \hbar (\tau^2)^{3/2} f(\tau) \s V^{1/2}>0$,
  so that the combination $\s^2 V^2$ always increases in time.
  (Therefore, $\s^2 V$ increases more strongly.) Afterwards, it is easy
  to  get $(i)$, by induction.
   Next, from eq.(21),
  $(\tau^2)^2(p_1^2 + p_2 ^2)$ increases,
   at least in the same manner as $\s^2 V^2$. Thus, we get $(ii)$.
   Now, from eq.$(24-b)$ and eq.(21),
  $$
 \dot{p_2} = -{\a \over 4V} \tau^2 (p_1^2 +p_2^2)
              -{3 \a \over 4 \tau^2} \s^2 V
              +\hbar (\tau^2)^{3/2}
               { {\partial f(\tau)} \over {\partial \tau^2}}
               V^{-1/2},
 \eqno{(24-b')}
  $$
  so that $\dot{p_2} < 0$ (note that
  ${ {\partial f(\tau)} \over {\partial \tau^2}} < 0 $). Furthermore,
  $|\dot{p_2}|> {3 \a \over 4 \tau^2} \s^2 V $, so that
 $\dot{p_2}$ decreases  faster than $-{\s^2 V \over \tau^2}$.
 (Note that, if $\tau^2 \rightarrow $finite, this implies
 a strong deceleration of $p_2$.)
 This fact excludes the behavior  $\tau^2 \rightarrow $finite, for, if so,
$p_2 \propto V {(\dot{\tau^2})^2 \over (\tau^2)^2   }$ (eq.$(24-a)$)
should tend to zero, which contradicts with the deceleration of
$p_2$. Therefore,
$\tau^2$ always behaves
   as $\tau^2 \downarrow 0$ or $\rightarrow \infty$. Thus, we get
   $(iii)$.  Next,
  we see  that $p_1 \dot{\tau}^1 + p_2 \dot{\tau}^2$
  ($= {\a (\tau^2)^2 \over V} (p_1^2 + p_2^2)$) (eqs.$(23-a)$, $(24-a)$)
   increases at least as $\s^2 V$,
   with the help of eq.(21). Finally,
    ${V \over (\tau^2)^2} ((\dot{\tau}^1)^2 +(\dot{\tau}^2)^2 )$
  ($\propto p_1 \dot{\tau^1} + p_2 \dot{\tau^2}$) increases at least
  as $\s^2 V$, so that
  ${1 \over (\tau^2)^2} ((\dot{\tau}^1)^2 +(\dot{\tau}^2)^2 )$
    increases at least as $\s^2 $. Thus, we get $(iv)$.

   The generic trajectories are the ones for which
 $\tau^1 \rightarrow $finite and
 $\tau^2 \rightarrow 0$, like Figures $2-a,b,c$. We can understand
  this behavior as follows: Suppose that $|\dot{\tau}^1|$ is at most
  comparable with $|\dot{\tau}^2|$. Then,
 from $(iv)$, we can make an estimation   as
$ {\dot{\tau}^2 \over \tau^2}$ $\sim - \s$, so that
   $\tau^2$ rapidly approaches to $0$ (faster than
   $\exp - \s t$ since $\s$ is increasing). Noting that $V^{-1}$ increases
    much slower than $\s$ ($(i)$), the combinations of
    the form  $(\tau^2)^n V^{-m}$ in $(23-a,b)$ become strong suppression
     factors. This  is compatible with  the assumption that
     $|\dot{\tau}^1|$ is not so large compared with $|\dot{\tau}^2|$.
      Therefore, in $(24-b)$, only the term proportional to
      $p_2^2$ on the R.H.S. dominates and determines the gross
      properties of the equation, which gives
      rise to  the universal behavior.

    There is  a special class of trajectories
  determined by
the initial condition $\tau^1=0$ (or in general, $\tau^1=n/2$ (integer))
 and $p_1 =0$.
 \footnote\star{
 The  remarks for  the last paragraph of $\S\S 3-b$ apply here, too.
  See the footnote there.
 }
  Due to  the property of $f(\tau^1, \tau^2)$ (eq.(18))
  with eqs.$(23-a,b)$,
  this implies that $\tau^1 \equiv 0$ (or $\equiv n/2$), $p_1 \equiv 0$,
   i.e. the trajectory of $(\tau^1, \tau^2)$ and $(p_1, p_2)$ form
  a line-segment on (or parallel to )
  the $\tau^2$-axis and $p_2$-axis, respectively. Combining   $(iv)$ with
   $p_1 \equiv 0$, we see that  $p_2 \dot{\tau}^2$
  always increases. It means that
   any  $(\tau^1, \tau^2)$-trajectory which is
  parallel to the $\tau^2$-axis has no turning point, and that $\tau^2$
  tends to $0$ or $\infty$, depending on  the initial condition.
  Furthermore, combining again $(iv)$ with  $p_1\equiv 0$ and
   $\tau^1$=constant, we see that
   ${\dot{\tau}^2 \over \tau^2}$$\sim \pm \s$, so that
   $\tau^2$ approaches rapidly to $\infty$ or $0$ (faster than
   $\exp \pm \s t$ since $\s$ is increasing).

 As is noted previously, our treatment is based on the adiabatic
 approximation. Thus, the results should always be taken with a caveat.
  In general, when instability is observed in the adiabatic treatment,
  it implies the unstable tendency of the system and it suggests
  the necessity of a further investigation beyond the adiabatic
  approximation, rather than just neglecting  the resultant instability.
   Furthermore, in the present case, there are good reasons to regard
   the unstable behavior as a real one. First, as investigated above,
   the universal
   asymptotic behavior of the generic trajectories implies that
   $\dot{\tau}^1 \rightarrow 0$,
   $\dot{\tau}^2 \rightarrow 0$ ( and  $\dot{V} \sim o(\s) $)
   although $p_2 \rightarrow -\infty $. This is because
   $\dot{\tau}^1 \propto { (\tau^2)^2 p_1 \over V}$,
   $\dot{\tau}^2 \propto { (\tau^2)^2 p_2 \over V}$,  and
   $\tau^2$ becomes a strong suppression (stronger than $\exp \s t$),
   while $1/V$ is at most $\sim \s$.
    Thus, the adiabatic treatment for
   $\tau^1$ and $\tau^2$ becomes better and better as
   $\tau^2 \rightarrow 0$:
${\dot{\omega}_A  / {\omega_A}^2} \sim
{\dot{\lambda}_A  /{ \lambda_A}^{3/2}} \sim
(\tau^2)^{3/2} \cdot { \dot{\tau}^2 \over (\tau^2 )^2 }
=(\tau^2)^{1/2} \cdot ({ \dot{\tau}^2 \over \tau^2})
 \rightarrow 0  $ (see eq.(5)).
 Furthermore,  $\dot V$ does not harm the adiabatic
    treatment because of the conformal
   invariance of the matter field,
   as has already been discussed previously
    ($\S\S2-(b)$, after
    eq.(8)). See Figures $3-a,b,c$.
(On the other hand, we should also note that
 the  special class of trajectories
characterized by $\tau^1 \equiv n/2$ (n:integer),
$\tau^2 \rightarrow \infty$, is not appropriate for the adiabatic
treatment: By $(iv)$, $\tau^2$ tends to infinity even stronger
than $\exp \s t$. However, because of the modular invariance,
the trajectories
for which $\tau^1\equiv$constant and  $\tau^2 \downarrow 0$ give
the good information of the class of these trajectories.)

Another support for the present result comes from the consideration
of the case of the negative cosmological constant without matter
field. It is straightforward to introduce the $\Lambda$-term [24]
(see eq.(16)):
$$
\eqalignno{
\a H & ={\a^2 (\tau^2)^2 \over {2V}}
         (p_1^2 + p_2^2)-{1\over 2}\a^2 \s^2 V
        - \a \Lambda V =0\ \ \ ,  & (21\Lambda) \cr
{\dot V}&=-\a \s V \ \ \ , & (22\Lambda-a) \cr
{\dot \s}&={\a \over 2} \s^2 + {\a (\tau^2)^2 \over {2V^2}}(p_1^2+ p_2^2)
                    + \Lambda  \ \ \ , & (22\Lambda-b) \cr
 \dot{\tau}^1 &={\a \over V} (\tau^2)^2 p_1 \ \ \ , & (23\Lambda-a) \cr
 \dot{p_1} &=0      \ \ \ , & (23\Lambda-b) \cr
 \dot{\tau}^2 &={\a \over V} (\tau^2)^2 p_2 \ \ \ ,& (24\Lambda-a) \cr
 \dot{p_2} &= -{\a \over V} \tau^2 (p_1^2 +p_2^2). & (24\Lambda-b) \cr
}
$$
 Here, $-\Lambda$ corresponds to the cosmological constant ($\Lambda >0$).
 Because of the negativity of the last term in eq.$(21\Lambda)$, the
 same kind of evolution for $(\tau^1, \tau^2)$
 as in the  case of the matter field is observed.
   (It is also notable that $(21 \Lambda)$-$(24 \Lambda-b)$ can be
   solved analytically [24].)
 It strongly suggests that the instability is independent of the
 adiabatic treatment. At the same time, we should  note the
 essential difference between our case and the case of the negative
  cosmological constant. Especially, the difference between
  $(23-b)$ and $(23\Lambda-b)$ is prominent. Furthermore,
   $\hbar (\tau^2)^{3/2}f(\tau) V^{-3/2}$ (which corresponds
   to $\Lambda$ comparing (21) with $(21\Lambda)$) depends on
   $(\tau^1, \tau^2)$ and $V$, which causes a highly non-trivial
   evolution.


\chapter{The effective action for the modular degrees of freedom}


\item{(a)} {\bf  Partition function}

 We have treated so far the back-reaction of the quantum field
  on the modular degrees of freedom, in the sense that the
   semiclassical Einstein equation, eq.(1),
has been solved, with $<T_{\a\b}>$ on the right-hand side being
calculated in the background spacetime. We can handle the same problem
 in a more systematic manner by the path-integral approach.
 The significance of this investigation is as follows:

    First, we know that we can derive eq.(1)  formally, by taking the
first variation of the phase w.r.t. $g_{\a \b}$
in the in-in path-integral expression for $g_{\a \b}$ and
$\psi$ [5][6][7].  However, when we discuss
the semiclassical gravity
in more detail, it is preferable to  take into account
the effects coming from the path-integral measure of $g_{\a\b}$.
 Since we cannot fix the measure in a reasonable manner,  we usually
do not discuss much about this effect. Fortunately, our model is
simple enough to investigate the measure to a great extent,
by making use of the techniques developed in string theories [15][16].

   Second, regarding this problem, we expect that the measure
in the original phase space, $\int [ dh_{ab} d\pi ^{ab}  dN dN_a ]$,
should reduce to  the standard canonical
measure in terms of the reduced  phase space variables,
$\int  [d\tau^A  dp_A][dV  d\sigma][dN]$, after gauge-fixing,
according to the general theory of the path integral for  the
1st-class constrained systems [25]. Analyzing this reduction process
in detail for the case of our model is   highly non-trivial and
helpful for deeper understanding of the path-integral approach for
quantum gravity [26].

   Third, furthermore, our model also becomes
a test candidate  for  another fundamental problem:
The validity of the minisuperspace approach in  quantum cosmology.
It is essential in our reduction procedure
that  the condition of  $N=$const on $\Sigma$ is compatible
with the equations of motion ($\S\S 3-b$). In the context of quantum
cosmology, it can  correspond to the minisuperspace approach: We often
impose the special form on metrics, which is compatible with
the equations of motion, and quantize them within this sub-class of
metrics, for tractability. Then, a fundamental question arises
as to whether this approximate treatment reflects
faithfully the main features of the full-quantized system.
The results may depend on which  space is chosen as
the starting whole  phase space,
viz. whether we start from the full phase space (full quantization)
or from its sub-space (minisuperspace quantization). In the former case,
it is expected that some extra factor emerges in the measure, since
in this case,
the condition $N=$const on $\Sigma$ should be treated as an extra
constraint, rather than just an ansatz. If so, this extra factor
can have some influence on the semiclassical evolution of the system.
The similar effect can arise from our assumption of the spatial
homogeneity of our torus model ($\S\S 2-a$).
Our model is suitable for the detailed analysis of this fundamental
problem.  In the present paper, however, we restrict ourselves to
the treatment {\it {\`a} la} minisuperspace models, which itself is one
consistent treatment.

      Fourth, when we need to investigate validity conditions of
the semiclassical treatment described by eq.(1), then, we have to study
the second variation of the effective action
$W[V_+, \tau_+;V_-, \tau_-]$ [7]. Thus, we need to estimate
$W[V_+, \tau_+;V_-, \tau_-]$ using  the in-in path-integral
formalism.

       We first discuss within the framework of the standard
 in-out  path-integral formalism [8]
 and later generalize it to the in-in formalism.
 In this subsection, we shall derive the expression for
  the partition function $Z$
in terms of the reduced phase space variables. In the next
subsection, we shall estimate the effective action for matter,
$W[V_+, \tau_+ ; V_-, \tau_-]$.

   The partition function in our case is given by
 $$
 \eqalign{
 Z &= {\cal N} \int [ dh_{ab} d\pi ^{ab}  dN dN_a ] [d\psi]
   \exp i\int ( \pi^{ab} {\dot h_{ab}} + p_\psi {\dot \psi}
           -N {\cal H} - N_a {\cal H}^a )   \cr
  &= {\cal N}\int [ dh_{ab}   d\pi ^{ab} dN dN_a ]
   \exp i\int ( \pi^{ab} {\dot h_{ab}}
           -N {\cal H} - N_a {\cal H}^a )\ \ \ , \cr
}
\eqno{(25)}
$$
where, in the last line, we understand that the matter
degrees $ \psi $ have been integrated out and  suitable
vacuum expectation values have  appeared in $\cal H$
 and ${\cal H}^a$
 (e.g. $T_{\a \b}n^{\a} n^{\b} \rightarrow
  <T_{\a \b}>n^{\a} n^{\b}$). (See the next subsection for
 more explicit discussions.)

Integrating over the multiplier $N_a$ is equivalent to
inserting $\delta ({\cal H}^a)$ and setting
$N_a$ to be an arbitrary value if needed.
\footnote\star{
    This situation is parallel to the case of QED. In the latter
    case, the term $A_0 div {\vec E}$ appears in the action.
    One  can set $A_0=0$ if needed, provided that
    $\delta (div {\vec E})$ is inserted in the integrand.
 }
 Let us set $N_a =0$:
$$
  Z= {\cal N} \int [ dh_{ab} d\pi ^{ab} dN ] \delta ({\cal H}^a)
   \exp i\int ( \pi^{ab} {\dot h_{ab}}
           -N {\cal H} )\ \ \ .
\eqno{(26)}
$$
The  action is invariant under the time-reparametrization and
$Diff(\Sigma)$ (the  diffeomorphism on the spatial surface $\Sigma$).
 Now, the gauge-fixing is needed to make this expression meaningful.
  The gauge-fixing condition for $Diff(\Sigma)$
  which is directly connected to our
  classical treatment in \S 3 is,
  $$
  h_{ab} - V \hh = 0\ \ \ ,
  \eqno(27)
  $$
  where $\hh$ is given in  $(2-b)$.

  At this stage, we need to fix our general attitude   for the treatment
  of our model.
  Any 2-dimensional metric $h_{ab}$ is conformally flat [15][16], and
  the conformal factor $V$ is a function of spatial coordinates (as well as
   a time parameter $t$) in general,
  $V=V(t, \xi^1, \xi^2)$. Here, furthermore, we set a further restriction
  to construct  a tractable model, which we have investigated in the previous
  sections: we restrict the class of spatial metrics $h_{ab}$
  to the one in  which  $V$ becomes spatially constant, $V=V(t)$.
  At the same time, the lapse function $N$ is restricted to
  $N=N(t)$. Both of these ansatz are compatible with
  the classical equations of motion.
  Such  restrictions on the class of the path-integral
  variables correspond to the minisuperspace models in quantum cosmology.
  (See \S 5 for more discussions on this point.)

  The treatment for the time-reparametrization invariance is
   well-investigated [27]. The final result is neat: Introducing
   the physical time $T=\int^t dt \  N(t)$, one computes
   a transition amplitude from time $0$ to time $T$. Then, integrate over
   the result with respect to $T$ [27]. Here, we shall not
   do it explicitly, since we are mainly interested in the
   semiclassical evolution of the system. We understand that we
   follow the above procedure whenever  needed.

   Then,
 $$
\eqalign{
   Z &= {\cal N}\int [dV dv'^a d^2\tau ][ dh_{ab} d\pi ^{ab} dN ]
   \delta (h_{ab}- V\hh) \Delta_{FP} \delta ((P_1^\dagger \pi)^a)
   \exp i S \cr
 &={\cal N} \int [dV dv'^a d^2\tau ][  d \tilde{\pi} ^{ab} d\sigma ]J
   [ dN ] \Delta_{FP\ |h_{ab}=V\hh} \delta ((P_1^\dagger {\tilde \pi})^a)
   \exp i S_{|h_{ab}=V\hh}\ \ \ ,\cr
}
$$
where $S = \int \pi^{ab} {\dot h_{ab}}
           -N {\cal H}$.
Note that $h_{ab}=V(t)\hh$ corresponds to choosing the York's time-slicing,
$K={\pi^a}_a /V =$const w.r.t. the spatial coordinates [21]
(see $\S\S 3-b$). Thus,
 only  the traceless part of $\pi^{ab}$,
${\tilde \pi}^{ab} = \pi^{ab}- {1\over 2} \pi h^{ab} = {\tilde K}^{ab} V$,
has remained in argument of the delta-function in the last line above.
Accordingly, the change of the integral variables
$\pi ^{ab} \rightarrow ({\tilde \pi}^{ab}, \sigma)$ has been performed and
$J$ is the Jacobian factor associated with this change.
Employing the  method in Appendix $B$, $J$ can be determined as follows:
A natural diffeo-invariant inner-product
\footnote\star{
   An appropriate power of $\alpha:=l_{\rm Planck}$ should be multiplied
   to formulae in order to adjust physical dimensions like eq.(A3).
    It is easy and not significant for the present discussions,
    so  we  omit  the factor.
}
 for $\delta \pi ^{ab}$ is
$(\delta \pi , \delta \pi )
= \int d^2 \xi \sqrt{h} h_{ac} h_{bd} \delta \pi ^{ab} \delta \pi ^{cd}$.
Substituting
$\pi ^{ab} = {\tilde \pi}^{ab} + {1 \over 2} h^{ab} \sigma V$, we get
$(\delta \pi, \delta \pi) =
(\delta {\tilde \pi}, \delta {\tilde \pi}) + {1 \over 2} V^2
(\delta \sigma, \delta \sigma)$, where
$(\delta \sigma, \delta \sigma)
= \int d^2 \xi \sqrt{ h } (\delta \sigma)^2$.
Thus,
$1= J \int d{\tilde \pi}^{ab} d\sigma \exp -(\delta \pi, \delta \pi)$,
 so that $J=V$ up to an unimportant numerical factor.

The Faddeev-Popov determinant $\Delta_{FP}$ in our case is equivalent to
the Jacobian associated with
the change of the integral variables from $h_{ab}$ to
$(V, v'^a, (\tau ^1, \tau ^2))$, where $v'^a \notin Ker P_1$.
Thus, we can employ the
method in Appendix $B$ again to determine $\Delta _{FP}$:
{}From eq. $(A1)$,
$$
\eqalign{
||\delta h_{ab} ||^2
 &=||\delta \phi h_{ab} + (P_1 v')_{ab}
 + {\cal T}_{Aab}\delta \tau ^A ||^2 \cr
 &= 4 (\delta \phi , \delta \phi) + (v', P_1^\dagger P_1 v')
  + ({\cal T}_A, {\cal T}_B)\delta \tau^A \delta \tau^B\ \ \ . \cr
}
  $$
Then,
$$
1 = \Delta_{FP} \int d\phi dv' d^2 \tau \exp -||\delta h||^2
= \Delta_{FP} (det' P_1^ \dagger P_1 )^{-1/2}
 det ^{-1/2}({\cal T}_A, {\cal T}_B)
 $$
 Thus,
 $$
 \Delta_{FP} =(det' P_1^\dagger P_1)^{1/2}
 det ^{1/2} ({\cal T}_A, {\cal T}_B)\ \ \ .
 $$
Thus,
$$
\eqalign{
Z  =  Vol_{ Diff_0} {\cal N}&
 \int [dV  d^2\tau ][  d \tilde{\pi} ^{ab} d\sigma ] [ dN ]\cr
  & \left(  det' P_1^\dagger P_1 \over det(\chi^\a, \chi^\b) \right)^{1/2}
 det ^{1/2} ({\cal T}_A, {\cal T}_B) V \d ((P_1^\dagger {\tilde \pi})^a)
   \exp i S_{|h_{ab}=V\hh}\ \ \ ,\cr
}
\eqno{(28)}
$$
where, $\{\chi^\a \}_{\a=1,2}$ is the  basis for $Ker P_1$, a space of
conformal Killing vectors.
\footnote\star{
   Any element in $Diff_0$
   (diffeomorphism on $\Sigma$ homotopic to 1)
   is associated with a vector $v^a$, which can be decomposed as
   $v^a= {v'}^a + \lambda _\a \chi^\a $, where ${v'}^a \notin Ker P_1$.
   Noting the argument in Appendix $A$,
   $\int [dv^a] =\int [dv'^a]\ d^2\lambda\  det^{1/2}(\chi^\a, \chi^\b)$,
    which   means $Vol_{Diff_0} = (\int [dv'^a])\cdot Vol_{Ker P_1}$.
     Thus, by   factorizing
     $(\int [dv'^a]) =  Vol_{Diff_0} / Vol_{Ker P_1}$
    from the path-integral, the factor $ det^{-1/2}(\chi^\a, \chi^\b)$
   appears. Here, the factor $(\int d^2\lambda)^{-1}$ is absorbed into the
   normalization ${\cal N}$.
   }

Let us investigate the factor $\d ((P_1^\dagger {\tilde \pi})^a)$.
According to eq. $(C1)$ in Appendix $C$ ($A=P_1^\dagger$,
${\vec x}={\tilde \pi}^{ab}$, $f({\vec x})=\exp iS$ and
$\{\Psi^A \}_{A=1,2}$ are the zero-modes for $P_1^\dagger$),
$$
\eqalign{
\int [d{\tilde \pi}^{ab} ]&\  \d ((P_1^\dagger {\tilde \pi})^a)
 \exp iS[h_{ab}=V\hh, {\tilde \pi}^{ab}, \sigma, N] \cr
& =\int [d^2 p]\
 { { det^{1/2} (\Psi^A, \Psi^B) } \over {det' P_1^\dagger} }
 \exp i \int dt (p_A \dot{\tau}^A
 + \sigma \dot{ V} - N {\cal H} )\ \ \ .\cr
 }
 \eqno{(29)}
 $$
Here, in the last line, the non-zero-mode components of
${\tilde \pi}^{ab}$ have been set to be zero according to
 the formula $(C1)$.
 This is equivalent to substituting
 ${\tilde \pi}^{ab}={\tilde K}^{ab}V= \sum_A p_A \Psi^{Aab}V $ into
  the action. Therefore, this is the path-integral version of
   the procedure of solving the momentum
  constraint in $\S\S 3-b$.

  Thus,
$$
\eqalign{
Z={\cal N}\int & [d\tau^A  dp_A][dV  d\sigma][dN]
 {{ {det'}^{1/2} P_1^\dagger P_1} \over { {det'}^{1/2} P_1 P_1^\dagger } }
 \cdot { {det ^{1/2} ({\cal T}_A, {\cal T}_B)}
  \over {det ^{1/2} (\chi^\a, \chi^\b) } }
  \cdot  det^{1/2} (\Psi^A, \Psi^B) \times \cr
& \times \exp i \int dt (p_A \dot{\tau}^A
 + \sigma \dot{ V} - N {\cal H} )\ \ \ .\cr
 }
 \eqno{(30)}
$$
Here, $det' P_1^\dagger= (det' P_1 P_1^\dagger)^{1/2}$ has been used.
\footnote\star{
    This equality can be shown by  estimating an  integral
    $I= \int dw'^{ab} \exp -(P_1^\dagger w', P_1^\dagger w')$  in two
    different manners (here, $w'^{ab}$ is symmetric,  traceless and
    $\notin Ker P_1^\dagger$):
    One way is $I=\int dw' \exp -(w', P_1 P_1^\dagger w')
    = (det' P_1 P_1^\dagger)^{-1/2}$, and the other way is
    $I= \int d(P_1^\dagger w') (det' P_1^\dagger)^{-1}$
    $\exp -(P_1^\dagger w', P_1^\dagger w')=(det' P_1^\dagger)^{-1}$.
    This change of the integral  variables in the latter
    estimation is valid since the space of
    the original variables ($w'^{ab}$) is isomorphic as a  vector space
    to  the space of the new variables
    ($(P_1^\dagger w')^a$)   by the map
    $P_1^\dagger$. See below.
 }

Now, we choose $\{ {\cal T}_A \}_{A=1,2}$ and $\{ \Psi^A \}_{A=1,2}$
 as $({\cal T}_A, \Psi^B)= \d_A^B$ (see $\S\S 3-b$), so that
 $det ^{1/2} ({\cal T}_A, {\cal T}_B)  det^{1/2} (\Psi^A, \Psi^B)
 =1$. For our case of a locally flat torus,
 $\{\chi^\a \}_{\a=1,2}$ can be chosen as
 $\chi^1_a = (1,0)$ and $\chi^2_a = (0,1)$, without inducing
 any critical point  as  vector fields. Then,
 $det ^{1/2} (\chi^\a, \chi^\b) =1$.

 Finally,
 ${det'}^{1/2} P_1^\dagger P_1$ and $ {det'}^{1/2} P_1 P_1^\dagger$
 should be estimated.
 The map $P_1$ is a map from a space of 2-vector fields to a space of
 2nd rank, symmetric and  traceless tensor fields,
 and the map $P_1^\dagger$
  is a map from the latter space to the former space.  Note that
  each  of the spaces can be represented as a 2-component  vector
  fields. Now, it is convenient to use the complex coordinates
  $(z, \bar{z})$, with respect to which both $P_1$ and
  $P_1^\dagger$ become diagonal [15].
   Let $z=x+iy$, ${\bar z}=x-iy$. Then, the  line element becomes
   ($e^\phi:=V$), $ds^2= e^\phi \hh d\xi^1 d\xi^2
   = e^\phi (dx^2+dy^2)= e^\phi dz\ d{\bar z}$, so that
   $h_{ab}= \pmatrix{0 & {1 \over 2} e^\phi \cr
                   {1 \over 2} e^\phi & 0 \cr     }_{(z, {\bar z})}$.
      (The suffix $(z, {\bar z})$ is  for the explicit indication of
       the coordinates employed.)
\footnote\S{
       We shall use  the following facts:
       $\partial :=  \partial_z = {1\over 2}(\partial_x -i \partial_y)$
        and  $\bar{\partial}
       :=  \partial_{\bar z} = {1\over 2}(\partial_x +i \partial_y)$;
       ${\vec v}= (v^1, v^2)_{(x,y)}
       = (v^1+iv^2, v^1-iv^2)_{(z, {\bar z})}$, i.e. $v^z=v^1+iv^2$,
       $v^{\bar z}=v^1-iv^2= \overline{ v^z}$ ($v^1, v^2 \in {\bf R}$);
       Let $T^{ab}$ be symmetric and traceless, and let its components
       in $(x,y)$-coordinates, $T^{11}$ etc., are real, then
       $(T^{ab})_{(z, {\bar z})}
       = diag \left( 2(T^{11}+iT^{12} ),
       \ 2(T^{11}-iT^{12}) \right)$, i.e.
       $T^{zz}=2(T^{11}+iT^{12})$,
       $T^{{\bar z} {\bar z}}=2(T^{11}-iT^{12})=\overline{T^{zz}}$ and
       the other components vanish; The Christoffel symbols become
       $\Gamma^z_{zz}=\partial \phi$,
       $\Gamma^{\bar z}_{{\bar z}{\bar z}}={\bar \partial} \phi
         =\overline{\Gamma^z_{zz} }$ and the others vanish.
}
The following arguments are valid
  for a general spatial metric $h_{ab}$ on a torus, so that we shall
  discuss in general terms. Only at the final stage (eq.(34) below),
  we set the condition that $\phi= \ln V $ = spatially constant.

 Now, both $P_1$ and $P_1^\dagger$ can be regarded as
 a map from a 2-component field to another 2-component filed:
 \nextline
  $P_1: {}^t(v^z, v^{\bar z}) \longmapsto
  {}^t \left( (P_1v)^{zz}, (P_1v)^{{\bar z}{\bar z}} \right)$,
  $P_1^\dagger: {}^t(w^{zz}, w^{{\bar z}{\bar z}})
  \longmapsto {}^t \left( (P_1^\dagger w)^z, (P_1^\dagger w)^{\bar z}
  \right) $, where $w^{ab}$ is a symmetric, traceless tensor field
   and ${}^t(\cdot, \cdot)$ indicates the transposition. In this sense,
   $P_1$ and $P_1^\dagger$ are represented as
$$
P_1 =\pmatrix{ 4e^{-\phi} \bar{\partial} &  0 \cr
                  0  &  4e^{-\phi} \partial       }\ , \ \
P_1^\dagger =\pmatrix{ -2e^{-2\phi}\partial e^{2\phi} &  0 \cr
                   0  &  -2e^{-2\phi}\bar{\partial} e^{2\phi}     }\ \ \ .
\eqno{(31)}
$$
Thus,
$P_1^\dagger P_1 : {}^t(v^z, v^{\bar z}) \longmapsto
         {}^t \left( (P_1^\dagger P_1 v)^z, (P_1^\dagger P_1)^{\bar z}
  \right)          $ is represented as
$$
P_1^\dagger P_1
=\pmatrix{ -8e^{-2\phi}\partial e^{\phi} \bar{\partial} &  0 \cr
                  0  & -8e^{-2\phi} \bar{\partial} e^{\phi} \partial }
=\pmatrix{ 2 \Delta + { {}^{(2)}\!  R}  & 0 \cr
             0          &  2 \Delta + {{}^{(2)}\!  R} }\ \ \ ,
\eqno{(32)}
$$
where $\Delta = -D_a D^a$ and $ {}^{(2)}\!  R$ are
 the Laplacian  and the scalar curvature, respectively,  defined by
 the covariant derivative ($D_a$) w.r.t. $e^\phi \hh$.  Similarly,
$P_1 P_1^\dagger : {}^t(w^{zz}, w^{{\bar z}{\bar z}})  \longmapsto
         {}^t \left( (P_1 P_1^\dagger w)^{zz},
                (P_1 P_1^\dagger w)^{{\bar z} {\bar z}} \right) $
 is represented as
$$
P_1 P_1^\dagger
=\pmatrix{ -8e^{-\phi}\bar{\partial} e^{-2\phi} \partial &  0 \cr
                  0  & -8e^{-\phi} \partial e^{-2\phi} \bar{\partial} }
=\pmatrix{ 2 \Delta -2\  {  {}^{(2)}\!  R}  & 0 \cr
             0          &  2 \Delta -2\  {{}^{(2)}\! R} }\ \ \ .
\eqno{(33)}
$$
 Therefore,
$$
det'^{1/2} P_1^\dagger P_1 = det'(2\Delta + { {}^{(2)}\! R})\ , \ \
det'^{1/2} P_1 P_1^\dagger = det'(2\Delta -2 { {}^{(2)}\! R})\ \ \ .
$$
 In our model of  locally flat tori ($\phi=\ln V=$ spatially constant),
 thus,
 $$
det'^{1/2} P_1^\dagger P_1 = det'^{1/2} P_1 P_1^\dagger
= det'(2\Delta)\ \ \ .
$$
Finally, we obtain
$$
Z={\cal N} \int [d\tau^A dp_A][dV d\sigma] [dN]
  \exp i \int (p_A \dot{\tau}^A + \sigma \dot{V} - N {\cal H})\ \ \ .
\eqno{(34)}
$$

  The integral region for $\tau^1, \tau^2$ should be understood as on
 the moduli space, ${\cal M}_{g=1}$:
As is indicated in eq.(28), $Diff_0 (\Sigma)$
(the  diffeomorphism group on $\Sigma$ homotopic to 1)
 has been factorized  from the path-integral. What is really  needed to
 be factorized  is the whole diffeomorphism group on $\Sigma$,
 $Diff(\Sigma)$.  Note that [15][16]
 $$
 \eqalign{
{\cal M}_g  &\simeq Riem (\Sigma)\big/Weyl \times Diff(\Sigma)
 \simeq \left(Riem (\Sigma)\big/Weyl \times Diff_0(\Sigma) \right)/MCG \cr
 &\simeq H_+ /PSL(2, {\bf Z}) \simeq D(H_+)/\sim \ \ \ .\cr
 }
 $$
 Here, $MCG := Diff(\Sigma)/ Diff_0 (\Sigma) $ is the
 mapping-class group for $\Sigma$, and
 $MCG \simeq PSL (2, {\bf Z})$ for $\Sigma \simeq T^2$
(i.e. a group of $2\times 2$ unimodular matrix with integer
elements,  modulo sign). $D(H_+)$ is the fundamental region in
$H_+$ (upper half-plane) w.r.t.  the action of $PSL (2, {\bf Z})$
(e.g. the Dirichlet region
$ D=\{ z \in H_+ \big| |Re z|\leq 1/2, |z|\geq 1   \}$)
and $/\sim$ indicates the identification
$(\tau^1, \tau^2)\sim -(\tau^1, \tau^2)$ on the boundary of
$D$ [15][16]. Thus,
the integral region for $(\tau^1, \tau^2)$ in eq.(34) should be
understood as over ${\cal M}_{g=1}$ rather than over $H_+$, considering
that we have factorized the volume of the mapping-class group
$MCG\simeq PSL (2, {\bf Z})$ as well as $Diff_0 (\Sigma)$ from  the
path-integral.

If we integrate out the momenta $p_A$ and $\sigma$ in eq.(34),
we get
$$
\eqalign{
Z= {\cal N} \int
\big[{ d\tau^A \over {(\tau^2)^2}}\big]
\big[{ dV \over \sqrt{V}}\big] [dN]
&  \exp i \int dt\  N(t) \big\{ {V \over  {2 \a (\tau^2)^2} }
 {1\over N^2}
 \left( (\dot{\tau}^1)^2 + (\dot{\tau}^1)^2 \right) \cr
& -{1\over {2 \a N^2 V}}\dot{V}^2
 +\hbar  (\tau^2)^{3/2}f(\tau) V^{-1/2} \big\} \ \ \ .\cr
}
\eqno{(35)}
$$
 Note that the kinetic term for $(\tau^1, \tau^2) $ in the action
 is proportional
 to ${\cal G}_{AB}\dot{\tau}^A \dot{\tau}^B$ and the last term
 in the action is proportional to the non-holomorphic
 Eisenstein series $G(\tau, 3/2)$ (see below eq.(18)). Thus,
 $Z$ is modular invariant since both the measure
 $d^2\tau \over {(\tau^2)^2}$ and the action are modular invariant.


\item{(b)} {\bf  Estimation of the functional determinant
 for the matter}

 Now, we estimate the path-integral for  the matter $\psi$  in eq.(25).
  Our aim is to obtain the effective action of the form
  $W[\phi, \tau^1(\cdot),\tau^2(\cdot)]$ by integrating out quantum
  fluctuations of the matter. Generalizing  the framework to the in-in
  formalism and getting $W[\phi_+, \tau_+; \phi_-, \tau_-]$, one can
  discuss the validity conditions for the semiclassical
  treatment [7], eq.(1).
At this stage, the peculiarity of the system including gravity is
prominent.
In the  standard treatment of a dissipative system, like a quantum
Brownian motion [17], the interaction between the sub-system and
 the environment is described by a weak, linear coupling. In our
case, however, there is no such interaction term between
gravity (analogous to the sub-system) and matter
(analogous to the environment). Rather, the interaction is
bilinear in $\psi$ and non-linear in $(\tau^1, \tau^2)$ and $\phi$,
 as is seen from  eq.(7). Thus,
it requires a new
treatment for a deeper analysis. Here, we should be  content with only a
rough estimation of the effect of the nonlinear coupling.
 We want to estimate the partition function for the matter,
 $$
 \eqalign{
 Z_\psi&=\int [d\psi] \exp -{1\over {2 \hbar} }\int \psi
 (-\tilde {\partial}^2 + {1\over 8} {\bar R})\psi \sqrt{\bar g }
 =Det^ {-1/2} \{ {V \over {2 \pi \hbar} }(-\tilde {\partial}^2
  +{1\over 8} {\bar  R})\} \cr
 &=\exp- {1\over  \hbar }{\tilde W}[\tau(\cdot)] \ \ .\cr
 }
 $$
 Here, `` $ \tilde{ \cdot}$ " denotes the
 Riemannian signature quantity. We calculate using  the metric
 ${\bar  g}_{\a \b}=(1, V \hh)$ with eq.$(2-b)$.
It is  difficult to estimate the above functional determinant
 exactly for a general function $(\tau^1(\cdot), \tau^2(\cdot))$ and
 $V(\cdot)$.
{}From the viewpoint of the quantum dissipative
 system, this difficulty comes  from the peculiarity of
 the interaction between gravity and matter.
  As discussed in the beginning of $\S\S2-b$,
  we treat the back-reaction problem in the sense
  that we investigate the modification of the background geometry
  due to matter, i.e. due to
  $<T_{\a\b}>$ calculated on the background spacetime.
   We have chosen as a background,
   a  flat spacetime. Thus, for  the
   lowest order approximation,
  we treat  $\tau^1$,  $\tau^2$ and $V$ as constants, so that we can set
  ${\bar R}=0$.
   This treatment corresponds to the lowest order estimation of the
   functional form of the effective potential in
   standard quantum field theory [28].

Thus, we need to estimate the determinant of the operator
$$
{\hat A}:= -{\a^2 V \over {2 \pi \hbar} }\tilde {\partial}^2
       = -{\a^2 V \over {2 \pi \hbar} }
       (\partial_0^2 + V^{-1} {\hat h}^{ab}\partial_a \partial_b)\ \ \ ,
$$
 where  $\hbar$ and $\a^2$ have  been inserted
  for the convenience of
  recovering a formula for pseudo-Riemannian signature.
   Now, we need to solve
the heat equation [28],
$$
\cases{
{\hat A}\rho=-{\partial \over \partial s} \rho\ \ \ ,\cr
lim_{s \downarrow 0} \rho(x,y,s)=\d^{(3)}(x-y)\ \ \ .\cr
}
$$
Here, $x:=(x^0=t, \xi^1, \xi^2)$.
Taking care of the periodicity in space, the solution is given by
$$
\eqalign{
\rho(x,x', s)=&\left( {\hbar \over {2 \a^2 V s}} \right)^{3/ 2}\times \cr
      & \times \sum_{n_1, n_2}
      \exp -{{\pi \hbar} \over {2 \a^2 V s} }
      \{ (x^0-x^{0'})^2+ V \hh (\xi-\xi'+n)^a (\xi-\xi'+n)^b  \},\cr
}
$$
especially,
$$
\rho(x,x, s)=\left( {\hbar \over {2 \a^2 V s}} \right)^{3/ 2}
      \sum_{n_1, n_2}\exp
      -{{\pi \hbar} \over {2 \a^2 s} }(n,n) \ \ \ ,
$$
where $(n,n):=\hh n^a n^b
    = {1\over \tau^2} (n_1^2+2\tau^1 n_1 n_2 +|\tau|^2 n_2^2)   $.
 Thus, the $\zeta$-function associated with $\hat A$ is [27],
$$
\eqalignno{
\zeta_A (z)
  &={1\over {\Gamma (z)}} \int_0^\infty ds s ^{z-1} Tr \rho (s) \cr
  &={\Omega \over {\Gamma (z)}} \sum_{n_1 n_2}
              \int _0^\infty ds  s^{z-1}
              \left({\hbar \over {2\a^2 V s}} \right)^{3/2}
             \exp -{ {\pi \hbar} \over {2\a^2 s}}(n,n)  \cr
        &=\left({\hbar \over {2\a^2 V}} \right)^{3/2}\Omega
        { { z \Gamma ({3\over 2} -z)} \over {\Gamma (z+1)} }
        \sum_{n_1 n_2}
        \left({ {\pi \hbar} \over {2\a^2 }}(n,n) \right)^{z-3/2}
                                          \ \ \ , & (36) \cr
}
$$
 where $\Omega=\int d^3x \sqrt{\bar g}$  and
 a transformation of variable $s$
 ($x:= { \pi \hbar \over {2\a^2 }}(n,n)s^{-1}$)
 has been done to get the formula in the last line from the middle line.
 Noting that
 ${d\over {dz}}|_{z=0}
 \left( {z\Gamma ({ {3\over 2} -z)} \over {\Gamma (z+1)} }
                C^{z-l}) \right)
                   ={{\sqrt \pi } \over 2} C^{-l}$ for
$\forall C$ when $C$ is independent of $z$, we get
$$
\zeta'_A (0)={\Omega \over 2\pi \a^3 V^{3/2}} \sum^{\ \ \prime}_{n_1 n_2}
                                   (n, n)^{-3/2}\ \ \ .
$$
Thus, we get
$$
\eqalignno{
{\tilde W}&={\hbar \over 2} \ln Det {\hat A}
                     =-{\hbar \over 2}\zeta'_A (0) \cr
        &=-{\hbar \Omega \over {4\pi \a^3 V^{3/2}}} (\tau^2)^{3/2}
    \sum_{n_1 n_2}^{\ \ \ \prime}
{1 \over (n_1^2+2\tau^1 n_1 n_2 +|\tau|^2 n_2^2 )^{3/2}} \ \ \ .& (37-a) \cr
}
$$
 To recover $W$ for the pseudo-Riemannian signature,
 we replace $\hbar \rightarrow i \hbar$, $\a \rightarrow i \a$
 (no change in   $\Omega$,
$d{\bar x}^0 dx^1 dx^2 dx^3 \leftrightarrow  dx^0 dx^1 dx^2 dx^3$).
 This replacement comes from the comparison between
 $W=i {\hbar \over 2} \ln Det \left(
       {i\a^2 \over {2 \pi \hbar} } (-{\partial}^2) \right)$ and
${\tilde W}= {\hbar \over 2} \ln Det \left(
       {\a^2 \over {2 \pi \hbar} } (-{\tilde \partial}^2) \right)$.
 Thus,
$$
W[\tau^1, \tau^2]
={\hbar \Omega \over {4\pi \a^3 V^{3/2}}} (\tau^2)^{3/2}
    \sum_{n_1 n_2}^{\ \ \ \prime}
{1 \over (n_1^2+2\tau^1  n_1 n_2 +|\tau|^2 n_2^2 )^{3/2}}\ \ \ .
\eqno(37-b)
$$

 Since we have used the expectation value of the energy-momentum
 tensor for the matter, $<T_{\a\b}>$, to couple with gravity
  (eq.$(13-b)$ or eq.(16)), we need to use the in-in path-integral
  formalism, rather than the standard in-out formalism [5][6][7].
   Then the matter part of the action (pseudo-Riemannian) (see eq.(7))
    should be reinterpreted as,
$$
\eqalign{
S_\psi
   &= -{1\over 2}\int_c
   ({\bar g}^{\a\b}\partial_\a \psi \partial_\b \psi
   +{1\over 8}{\bar R}\psi^2)
           \sqrt{-{\bar g}} \cr
   &= -{1\over 2}\int_+
   ({\bar  g}^{\a\b}\partial_\a \psi \partial_\b \psi
    +{1\over 8}{\bar R}\psi^2 )      \sqrt{-{\bar g}}
    +{1\over 2}\int_-
    ( {\bar g}^{\a\b}\partial_\a \psi \partial_\b \psi
      +{1\over 8}{\bar R}\psi^2 )
           \sqrt{-{\bar g}}\ \ \ , \cr
}
$$
where ``\ $c$\ " stands for the closed-time contour and
``$\  +\ $" and  ``$\ -\ $" stand for, respectively,
the $+$-branch and  the $-$-branch of the time-contour.
Then,
$$
\eqalign{
S_\psi&=\int_+ \sqrt{-{\bar g}_{{}_+} } \psi
             {1\over 2\hbar}
             (\partial^2  -{1\over 8}{\bar R})\psi
        -\int_- \sqrt{-{\bar g}_{{}_-} }
   \psi {1\over 2\hbar}(\partial^2  -{1\over 8}{\bar R})\psi \cr
      &=\int (\psi_+\   \psi_-)
\pmatrix{
{1\over {2\hbar}}(\partial_{{}_+}^2 -{1\over 8}{\bar R})
  \sqrt{-{\bar g}_{{}_+}} & 0 \cr
  0 & -{1\over {2\hbar}}
       (\partial_{{}_-}^2 -{1\over 8}{\bar R} )
       \sqrt{-{\bar g}_{{}_-}} \cr}
           \pmatrix{\psi_+ \cr
                 \psi_- \cr}\ \ \ .
}
$$
Since $+$ and $-$ components are separated completely, it is enough to
look at  only the $+$-sector (or $-$-sector).

 Now, let  us investigate the effective action, \ \
$S[ \tau^1, \tau^2, V; N]= S_g[ \tau^1, \tau^2, V; N] $
$+ W[ \tau^1, \tau^2, V; N] $, where
$S_g[\phi, \tau^1, \tau^2; N]$ is the reduced action for gravity in
terms of the configuration variables and
$W[ \tau^1, \tau^2, V; N] $  is  given by eq.$(37-b)$.
The effective action
$S[ \tau^1, \tau^2, V; N]$ is what has appeared in the exponent in
eq.(35).
  It should be noted that
 the first variations of $S[\phi, \tau^1, \tau^2]$ w.r.t.
 $N$, $V$ and $(\tau^1, \tau^2)$ reproduce exactly eqs.$(21-24)$.
  This result shows the following two points.

     First,  our  approximation for the estimation of
  $Det^{-1/2} \left( {1\over \hbar}(-\tilde{ \partial}^2
  + {1\over 8} {\bar R}) \right)$,  treating $\tau^1, \tau^2$ and V
  as if they were constants so that ${\bar R} =0$, corresponds
  to the approximation used to solve the semiclassical
  Einstein equation, eq.(1). Namely, $<T_{\a\b}>$, calculated
  on a flat background, is used in eq.(1) to estimate
  the deviation from the original background geometry.
   As is discussed at the beginning of $\S\S 2-b$,
  the latter approximation has been implemented for
  the tractability of the problem, at the expense of
  the self-consistency  of eq.(1). Such an approximation is what is
  usually  meant by the term ``back-reaction",
  and this may be the best we can do in practice.

     Second, regarding the path integral expressions
      in the Lagrangian formalism, like eq.(35):
 We can reproduce eq.(1) (or equivalently,
 eqs.(22-24)) from the phase part $S_g+W$ in the partition function
 $Z$ with the matter part integrated, and without
 taking care of the contributions from the measure for $V$ and
 $(\tau^1, \tau^2)$ (see eq.(35)). However, we now know explicitly the
 non-trivial path-integral measure for $V$ and $(\tau^1, \tau^2)$ as
 is shown in eq.(35).
There should be $O(\hbar)$ correction to eq.(1) coming from the
path-integral measure for $g_{\a\b}$ and this correction will cause
a non-trivial correction to the dynamics of $g_{\a\b}$. We shall
come back to this point in the next section.


\chapter{Discussions}

In this paper,  we have  investigated  the
semiclassical dynamics of the topological degrees of freedom,
 $(\tau^1, \tau^2)$, which has
 been seldom discussed so far. By reducing the spacetime dimension
 to 3, we could concentrate on the study of a finite number of
 topological
 modes and we could describe the back-reaction effect from matter
 to topological modes, explicitly. We observed a non-trivial
 dynamics caused by the back-reaction.
 The back-reaction makes the toroidal universe unstable:
 The shape of the torus becomes
     thinner and thinner, while its total 2-volume  becomes
    smaller and smaller. These are universal behaviors of the system
     independent of the initial conditions, which is justified
     by the asymptotic analysis of the set of dynamical equations.
   This observation implies
 the importance of the investigation of topological aspects for
 a deeper understanding of quantum gravity. Moreover,
 we could fix the path-integral measure for $(\tau^1, \tau^2)$ and $V$
 and observe that the partition function is expressed in terms of the
 canonical variables for
  the reduced phase space with the standard Liouville measure.

Let us note a few points regarding   the path-integral measure.

We obtained
  the path-integral expression on the reduced phase space
with the Liouville measure (eq.(34)), while  the path integral
 on the configuration space requires a non-trivial
measure (eq.(35)). Indeed, the combination
${ d\tau^A \over  (\tau^2)^2}$ is essential to make the partition function
modular invariant. In our model, the semiclassical Einstein equation,
eq.(1),  corresponds to eqs.$(21)-(24)$,  and they are derived
from the variation of the exponent in eq.(34) or eq.(35).
It means that, from the viewpoint of the Lagrangian formalism,
 the semiclassical Einstein equation is derived from the
 variation of the phase part in the partition function, with
 the measure factor untouched. Thus, the measure factor gives
 the $O(\hbar)$-correction to eq.(1). In our model,
 the term  $\int dt N(t) \hbar (2\ln \tau^2 + {1/2}\ln V)$
 can be added to the action as a correction. (Note the time
  reparametrization invariance implied  in eq.(34).) Then,  it is
  a non-trivial question worth while to investigate which
  is better as the semiclassical description,
  the semiclassical Einstein equation in terms of the canonical
   variables (eqs.(21)-(24) in our case), or the same in terms
   of the configuration variables with suitable corrections originating
   from the measure. If we perform the path integral exactly,
   both the canonical and the Lagrangian formalisms will give
    equivalent results,
   but they will  not be equivalent  within the accuracy
   of the semiclassical approximation.

 Another important problem is linked  with the validity of the
 minisuperspace treatment. We have investigated the homogeneous
 model, which is equivalent to assuming  $N=N(t)$, $V=V(t)$ (see
 the discussion in $\S\S 4-a$, below eq.(27)). We can set this ansatz
 since it is compatible with the dynamics. This treatment corresponds
 to the minisuperspace approach in quantum cosmology.
 Though such a  treatment is completely
 self-consistent, it is important to question
 to what extent such a treatment reflects
   the original full quantum theory faithfully.
   From the viewpoint of the original
   full system, the restrictions are regarded as extra constraints on
   the phase space.  These constraints
    can modify the path integral measure for
   the reduced variables (minisuperspace variables).
   Since this problem is a fundamental one, it should be
   investigated separately. Our model may  be
   a good test candidate to investigate this point in detail.


\ack{
The author thanks R.~Balasubramanian, T.~Ghosh, S.D.~Mohanty
 and V.~Sahni for helpful discussions.
 This work has been financially supported
 in part by the Honda Fellowship of
 the Japan Association for Mathematical Sciences, and the
 Yukawa Memorial Foundation, Japan.
}

\appendix

{\bf A. Brief summary on the moduli space}

We give here a concise summary on the moduli space just for
fixing  the terminology and notations  used in \S 3 and \S4.
See e.g. [15], [16] for more detailed information.

 Let $\Sigma$ be a
  2-dimensional, compact, closed, orientable manifold  with genus $g$.
 The moduli space ${\cal M}_g$ of $\Sigma$ is defined as
 $
{\cal M}_g \simeq  Riem (\Sigma)\big/Weyl \times Diff(\Sigma)
$, where
  $Riem (\Sigma)$ is a space of all Riemannian
 metrics on $\Sigma$, $Weyl$ is for    the
 Weyl  group  and $Diff(\Sigma)$ is for
  the diffeomorphism group on $\Sigma$. The universal covering
  space of ${\cal M}_g$ is called  the Teichm\"uller space.
  Now, the   tangent space of ${\cal M}_g$, $T({\cal M}_g)$, can be
  investigated as follows:
 Any variation of the spatial metric $\d h_{ab} \in  T(Riem(\Sigma))$,
 can be decomposed into the trace part and the traceless part,
 the latter is furthermore  decomposed into  the
 diffeomorphism $\d_D h_{ab}$, and the moduli deformation
 $\d_M h_{ab}$;
 $$
\eqalignno{
\d h_{ab}&= \d_W h_{ab} + \d_D h_{ab} +\d_M h_{ab} \cr
         &=\d_W h_{ab} + (P_1v)_{ab} + {\cal T}_{Aab}\d \tau^A\ \ \ ,
                                                   &(A1) \cr
}
$$
where
$$
\eqalignno{
\d_W h_{ab}&=\d \phi h_{ab}\ \ \ {\rm for}\ \ \ {\exists \d \phi}\ \ \ ,
                                   &(A2-a) \cr
(P_1v)_{ab}&=D_a v_b +  D_b v_a - D_c v^c h_{ab}
                           \ \ \ {\rm for}\ \  \ {\exists v^a} \ \ \ ,
                                             &(A2-b) \cr
 {\cal T}_{Aab}&={\partial h_{ab} \over \partial \tau^A}
  -{1\over 2}h^{cd}{\partial h_{cd} \over \partial \tau^A }h_{ab}\ \ \ .
                                                   &(A2-c) \cr
}
$$
Here $\{\tau^A \}_{A=1,..., dim_{\bf R}{\cal M}_g }$ are the Teichm\"uller
parameters  specifying  a point in ${\cal M}_g$.
 A  natural inner-product on $T(Riem(\Sigma))$ is introduced as
$$
(A,B):={1\over \a^2}\int_{\Sigma} d^2x \sqrt{h}\  h^{ac}h^{bd}
A_{ab} B_{cd}
  \ \ \ {\rm for }\ \ \ \forall A_{ab},\
  \forall B_{ab} \in T(Riem(\Sigma)) \ \ \ ,
 \eqno(A3)
$$
where $\a$ is the Planck length, inserted to adjust
the physical dimension.
Then,  the tangent space of the moduli space,
$T({\cal M}_g)$,
can be characterized by the  set of all symmetric, traceless (covariant)
 tensors  which
are  perpendicular to $T(Diff(\Sigma))$ w.r.t. the inner-product
$(A3)$, the latter condition being equivalent to the condition
 $$
 ({P_1}^\dagger w)^a =-2D_bw^{ab}=0\ \ \ ,
 \eqno{(A4)}
 $$
for $w\in T^* ({\cal M}_g)$.
 Thus, $dim_{\bf R}{\cal M}_g=dim_{\bf R}T^* ({\cal M}_g)
     =dim_{\bf R}Ker{P_1}^\dagger$, which is known as
     $=0$, $=2$, and $=6g-6$ for $g=0$, $g =1$, and $g \geq 2$,
      respectively.
It is also known that
 $dim_{\bf R} Ker{P_1}  - dim_{\bf R} Ker{P_1}^{\dagger} = 6-6g$
 (Riemann-Roch theorem).
 For the  case of a torus ($g=1$), then,
$dim_{\bf R}{\cal M}=2$ and  $dim_{\bf R}Ker{P_1}=2$. Thus,
       two Teichm\"uller parameters $(\tau^1, \tau^2)$ are needed to
      describe the modular deformations
      $\d_M h_{ab} \in T({\cal M}_{g=1})$, and two independent
      vectors $\{ \chi^\a \}_{\a=1,2}$ are needed as the basis of
      $Ker {P_1}$.

 Let
 $\{ { {\cal T} _A}_{ab} \}_{A=1,2, \cdots , dim_{\bf R} {\cal M}_g }\ $
   be the basis of
$T ({\cal M}_{g})$,  and
 $\{ {\Psi ^A}^{ab} \}_{A=1,2, \cdots dim_{\bf R} {\cal M}_g}$
 be the basis of $T^* ({\cal M}_{g})$. They can be chosen to satisfy
 $( \Psi ^A, {\cal T}_B) ={\d^A}_B$.
 Then,
they define a metric on
${\cal M}_{g=1}$ (the Weil-Peterson metric),
induced from the inner-product eq.$(A3)$ on
$T(Riem(\Sigma))$,
$$
\eqalign{
{\cal G}_{AB} &=({\cal T}_A, {\cal T}_B)\ \ \ , \cr
{\cal G}^{AB} &=( \Psi^A, \Psi^B)
             ={\rm inverse\ matrix\  of\  }{\cal G}_{AB}\ \ \ . \cr
}
\eqno(A5)
$$

{\bf B. The Jacobian associated with a change of integral variables.}

Let us note a convenient method to specify the Jacobian
 associated with a change of integral variables.(See e.g. [15].)

  If a line element $ds$ is given on a space of integral variables
   ($X^A$, $A=1, 2, \cdots, n $) as
   $ds^2 = G_{AB} dX^A dX^B =: (dX, dX) $, then
   $d^nX \sqrt{det\ G}$ is a natural integral measure, where
   $\sqrt{det\ G }$  takes care of the Jacobian factor.
    Suppose we  change the variables from $X^A$ to $X^{A'}$,
     then $d^nX'\sqrt{det\ G'}$ is the corresponding
     integral measure for the new variables.
     Now, a convenient way to find out the expression for
     $\sqrt{det\ G'}$
      is
  \item{1)} Express $\d X^A$ in terms of $\d X^{A'}$,
     $\d X^A = {\partial X^A \over \partial X^{A'}  }\d X^{A'}$.
 \item{2)} Then express $(\d X, \d X)$ in terms of $\d X'$,
      $(\d X, \d X)= { \partial X^A \over \partial X^{A'} }
              { \partial X^B \over \partial X^{B'} }
                    (\d X^{A'}, \d X^{B'})   $.
  (This should be equivalent to $G_{A' B'} \d X^{A'} \d X^{B'}$.)
 \item{3)} Then determine the Jacobian $J$ by setting
 $ 1=J \int d^n \d X' \exp -(\d X, \d X) $, since this should be
  equivalent to
  $1= J({  det\  G'  / \pi  })^{-1/2} $.
  (The factor $\pi$ is usually  unimportant and omitted.)

\vskip .5cm

{\bf C. A formula for the delta-function.}

Let us derive a formula which modifies an integral
including  $\d (A {\vec x})$ into  a
 more practical form. Here, $A$ is a linear operator
 possibly with zero-modes.

Let us consider an integral,
$I=\int d {\vec x}\  \d (A {\vec x} ) f({\vec x})$.
Let $\{ \Psi^A \}$ ($A=1, 2, \cdots, m=dim Ker A $) be the zero-modes
for $A$. Then, any element  $\vec x$ of a vector space $\cal V$
can be decomposed as
${\vec x}= {\vec X} + \sum_A p_A {\vec \Psi}^A$, where
${\vec X} \in {\cal V} \big/ Ker A$. Now we change the integral variables
from $\vec x$ to $({\vec X}, \ p_A)$. Then,
$({\vec x}, {\vec x})=({\vec X}, {\vec X})
+ ({\vec \Psi}^A, {\vec \Psi}^B)p_A p_B$, where $(\cdot, \cdot)$
 is a suitable inner-product, which is assumed to be given.
Thus, according to Appendix $B$, the associated Jacobian $J$ becomes
$J= det^{1/2} (\Psi^A, \Psi^B)$. Thus,
$$
\eqalign{
I&= \int d{\vec X} d{\vec  p} \   det^{1/2} (\Psi^A, \Psi^B)
 \d (A {\vec X}) f({\vec X},\  {\vec p}) \cr
 &=\int d{\vec  p}\  det^{1/2} (\Psi^A, \Psi^B) (det' A)^{-1}
  f({\vec X}={\vec 0},\ {\vec  p} )\ \ \ , \cr
}
 $$
 where an equality,
 $\d (A {\vec X})=(det' A)^{-1} \d ( {\vec X}) $ when
 $\vec X \in {\cal V} \big/ Ker A $,  has been  used in the last line.
 (This equality can be shown easily by the variable change
 from $\vec X$  to ${\vec Y}=A {\vec X}$). Therefore, we have
 obtained a formula
$$
\int d {\vec x} \d (A {\vec x} ) f({\vec x})
=\int d{\vec  p}\  det^{1/2} (\Psi^A, \Psi^B) (det' A)^{-1}
  f({\vec X}={\vec 0},\  {\vec  p})\ \ \ .
\eqno(C1)
$$


\REF\one{See e.g., J.A.~Wheeler, Ann. Phys. {\bf 2} (1957), 604;
 R.P.~Geroch,  J. Math. Phys. {\bf 8} (1967), 782;
  R.D.~Sorkin, Phys. Rev. {\bf D33} (1986), 978;
 M.Visser, Phys. Rev. {\bf D41} (1990), 1116;
 G.W.~Gibbons and J.B.~Hartle, Phys. Rev. {\bf D42} (1990), 2458.}
\REF\two{ N.D.~Birrell and P.C.W.~Davies,
{\sl  Quantum Fields in Curved Space} (Cambridge
University Press, 1982).}
\REF\three{ S.A.~Fulling, {\sl Aspects of Quantum Field
Theory in Curved Space-Time} (Cambridge University Press, 1989). }
\REF\four{See e.g., T.~Padmanabhan and T.P.~Singh,
 Ann. Phys. {\bf 221} (1992), 217.}
\REF\five{R.D.~Jordan,  Phys. Rev. {\bf D33} (1986), 444.}
\REF\six{E.~Calzetta and B.L.~Hu,  Phys. Rev. {\bf D28} (1987)
, 495.}
\REF\seven{M.~Seriu and T.P.~Singh,  Phys. Rev. {\bf D50} (1994), 6165.}
\REF\eight{J.~Schwinger,  J. Math. Phys. {\bf 2} (1961), 407.}
\REF\nine{L.H.~Ford,  Ann. Phys. {\bf 144} (1982), 238.}
\REF\ten{T.P.~Singh and T.~Padmanabhan,  Ann. Phys. {\bf 196}
              (1989), 296.}
\REF\eleven{E.~Martinec, Phys. Rev. {\bf D30} (1984), 1198.}
\REF\twelve{V.~Moncrief, J. Math. Phys. {\bf 30} (1989), 2907.}
\REF\thirteen{A.~Hosoya and K.~Nakao, Prog. Theo. Phys.
{\bf 84} (1990), 739.}
\REF\fourteen{M.~Seriu, Phys. Lett. {\bf B319} (1993), 74.}
\REF\fifteen{B.~Hatfield, {\sl Quantum Field Theory of
Point Particles and Strings} (Addison -Wesley, 1992).}
\REF\sixteen{D.~L\"ust and S.~Theisen,
{\sl Lectures on String Theory} (Springer-Verlag, 1989).}
\REF\seventeen{A.O.~Caldeira and A.J.~Leggett, Physica {\bf 121A}
 (1983), 587.}
\REF\eighteen{See e.g., I.~Chavel,
{\sl Eigenvalues in Riemannian Geometry} (Academic Press 1984),
Chapter $XI-2$.  }
\REF\nineteen{For a systematic investigation of the Casimir energy
including more general cases,
see J.S.~Dowker, J. Math. Phys. {\bf 28} (1987), 33;
  Phys. Rev. {\bf D40} (1989), 1938. }
\REF\twenty{R.M.~Wald, {\sl General Relativity}
(University of Chicago Press, 1984).}
\REF\twentyone{J.W.~York, J. Math. Phys. {\bf 14} (1973), 456.}
\REF\twentytwo{D.~Zagier, in
 M. Waldschmidt, P. Moussa, J.-M. Luck and
 C. Itzykson (Ed.), {\sl From Number Theory to Physics}, Springer-Verlag,
 Berlin Heidelberg, 1992.}
\REF\twentythree{R.~Abraham, J.E.~Marsden, T.~Ratiu,
{\sl Manifolds, Tensor Analysis, and Applications } (2nd ed.)
 Springer-Verlag, 1988.}
\REF\twentyfour{
Y.~Fujiwara and J.~Soda, Prog.Theo.Phys. {\bf 83} (1990), 733;
K.~Ezawa, Phys. Rev. {\bf D49} (1994), 5211;
S.~Carlip and J.E.~Nelson, gr-qc/9411031.}
\REF\twentyfive{ M.~Henneaux and C.~Teitelboim,
{\sl Quantization of Gauge Systems },
 Princeton University Press, Princeton, 1992, Chapter 16.}
\REF\twentysix{The same problem is analysed for the case of $g \geq 2$
 in S.~Carlip, Class. Quantum Grav. {\bf 12} (1995), 2201.}
\REF\twentyseven{ J.J.~Halliwell, Phys. Rev. {\bf  D38} (1988), 2468.}
\REF\twentyeight{ See  e.g., P.~Ramond, {\sl Field Theory: A Modern Primer}
 (2nd ed.), Addison-Wesley,  1989.}

\refout

\endpage


\nopagenumbers

\centerline{Figure Captions}

\item{Figure $1-a$:}
The plot of the function $f(\tau^1, \tau^2)$ for the range
$\tau^1:0-1$ and $\tau^2:0.5-0.8$. The infinite summation has been
truncated at $-200$ and $200$.

\item{Figure $1-b$:}
The contour plot of $f(\tau^1, \tau^2)$, with the same range and
the truncation points as in Figure $1-a$. The lines indicate the values
 (from bottom to top) 30, 28, 26, 24, 22, 20, 15  and 5.

 \item{Figure $2-a$:}
 The trajectory of $(\tau^1, \tau^2)$ determined by eqs.(21)-(24).
 The infinite summation in the  definition of  $f(\tau)$ has
 been truncated at $-200$ and $200$. $\hbar$ and $\alpha $ have
  been set unity.
 The initial conditions are $\tau^1=0.500$, $\tau^2=0.500$, $p_1=1.000$,
  $p_2=1.620$, $V=1.000$ and $\sigma=0.000$.
  Points $A-F$ and $Z$
   indicate typical points on the trajectory. $A$: the initial point,
    $B$: a point near the  turning point ($p_2=0$),
    $C-F$: the  points for which
    $t$ is near integer, $Z$: the end point of the calculation.
     $A:(0.000, 0.500)$ at $t=0.000$,
    $B:(0.727, 0.888)$ at $t=0.675$, $C:(1.171, 0.760)$ at $t=1.003$,
    $D:(1.551, 0.161)$ at $t=2.057$,
    $E:(1.553, 0.031)$ at $t=2.953$,
    $F:(1.549, 4.8\times 10^{-3})$ at $t=4.043$
    and $Z:(1.549, 1.0 \times 10^{-3})$ at $t=4.983$.

 \item{Figure $2-b$:}
 The trajectory of $(p_1, p_2)$ determined by eqs.(21)-(24).
 The initial conditions are the same as
 in Figure $2-a$. $A:(1.000, 1.620)$ at $t=0.000$,
 $B:(0.949, -3.6\times 10^{-3})$
 at $t=0.675$, $C:(0.949, -0.603)$ at $t=1.003$, \nextline
 $D:(0.811, -5.301)$ at $t=2.057$,
    $E:(-4.960, -29.22)$ at $t=2.953$,
    $F:(-9.310, -175.3)$ at $t=4.043$,
  and $Z:(-12.59, -820.4)$ at $t=4.983$.

 \item{Figure $2-c$:}
 The trajectory of $(V, \sigma)$ determined by eqs.(21)-(24).
 The initial conditions are the same as in Figure $2-a$.
 $A:(1.000, 0.000)$ at $t=0.000$, $B:(1.089, -0.235)$
 at $t=0.675$, $C:(1.193, -0.320)$ at $t=1.003$,
 $D:(1.825, -0.452)$ at $t=2.057$,
 $E:(2.774, -0.474)$ at $t=2.953$, \nextline
 $F:(4.540, -0.423)$ at $t=4.043$
 and $Z:(6.597, -0.372)$ at $t=4.983$.

 \item{Figure $3-a$:}
 The value of $\dot{\tau}^1$ during the evolution shown in
 Figure $2-a,b,c$. $A: 0.500$ at $t=0.000$,
 $B: 1.500$ at $t=0.675$,
 $C: 1.100$ at $t=1.003$,
$D: 4.182 \times 10^{-2}$ at $t=2.057$,
$E: -9.657 \times 10^{-3}$ at $t=2.953$,
 $F: -4.290 \times 10^{-4}$ at $t=4.043$
  and $Z :-2.519 \times 10^{-5}$ at $t=4.983$.

 \item{Figure $3-b$:}
 The value of $\dot{\tau}^2$ during the evolution shown in
 Figure $2-a,b,c$. $A: 0.810$ at $t=0.000$,
 $B: -5.678 \times 10^{-3}$ at $t=0.675$,
 $C: -0.697$ at $t=1.003$,
$D: -0.273$ at $t=2.057$,
$E: -5.688 \times 10^{-2}$ at $t=2.953$,
 $F: -8.078 \times 10^{-3}$ at $t=4.043$
  and $Z :-1.641 \times 10^{-3}$ at $t=4.983$.

 \item{Figure $3-c$:}
 The value of $\dot{V}$ during the evolution shown in
 Figure $2-a,b,c$. $A: 0.000$ at $t=0.000$,
 $B: 0.257$ at $t=0.675$,
 $C: 0.382$ at $t=1.003$,
$D: 0.825$ at $t=2.057$,
$E: 1.315$ at $t=2.953$,
 $F: 1.922$ at $t=4.043$
  and $Z : 2.454$ at $t=4.983$.

\end